\documentclass[a4paper,twoside,twocolumn,english,aps,superscriptaddress,showpacs]{revtex4}
\usepackage{times}
\usepackage[T1]{fontenc}
\usepackage[latin1]{inputenc}
\usepackage{amsmath}
\usepackage{graphicx}
\usepackage{amssymb}

\makeatletter

\newcommand{\lyxdot}{.}

\usepackage{babel}
\makeatother
\begin{document}

\title{Composite fermionization of 1-D Bose-Bose mixtures}

\author{Sascha Zöllner}

\email{sascha.zoellner@pci.uni-heidelberg.de}

\affiliation{Theoretische Chemie, Universit\"{a}t Heidelberg, INF 229, 69120
Heidelberg, Germany}

\author{Hans-Dieter Meyer}

\affiliation{Theoretische Chemie, Universit\"{a}t Heidelberg, INF 229, 69120
Heidelberg, Germany}

\author{Peter Schmelcher}

\email{peter.schmelcher@pci.uni-heidelberg.de}

\affiliation{Theoretische Chemie, Universit\"{a}t Heidelberg, INF 229, 69120
Heidelberg, Germany}

\affiliation{Physikalisches Institut, Universit\"{a}t Heidelberg, Philosophenweg
12, 69120 Heidelberg, Germany}

\begin{abstract}
We study the ground states of one-dimensional Bose-Bose mixtures under
harmonic confinement. As we vary the inter-species coupling strength
up to the limit of infinite repulsion, we observe a generalized, \emph{composite-fermionization}
crossover. The initially coexisting phases demix as a whole (for weak
intra-species interactions) and separate on an atomic level (for strong
intra-species repulsion). By symmetry, the two components end up with
strongly overlapping profiles, albeit sensitive to symmetry-breaking
perturbations. Different pathways emerge in case the two components
have different atom numbers, different intra-species interactions,
or different masses and/or trap frequencies. 
\end{abstract}

\date{\today}

\pacs{67.85.-d, 67.60.Bc, 03.75.Mn}

\maketitle

\section{Introduction}

The availability of cold atoms has made it possible to realize many
fundamental quantum systems. Building on the seminal realization of
Bose-Einstein condensation \cite{pethick,pitaevskii}, mixtures composed
of, say, two different atomic species have come into the research
focus. Aside from Bose-Fermi \cite{truscott01,hadzibabic02}  or
Fermi-Fermi mixtures \cite{taglieber08}, whose potential for studying
phenomena as diverse as impurity effects or superconductivity has
been recognized more recently, two-component bosonic mixtures have
received much experimental \cite{myatt97,hall98,maddaloni00,modugno02,catani08}
and theoretical attention (see \cite{cazalilla03,li03,shchesnovich04,alon06,mishra07,roscilde07,nho07,kleine08}
and Refs. therein). The interplay between intra- and inter-species
forces gives rise to many effects not accessible with single-component
Bose gases, including phase separation and modified superfluid-insulator
transitions \cite{alon06,mishra07}, quantum emulsions \cite{roscilde07},
and spin-charge separation \cite{kleine08}.

Most studies so far have focused on the regime of relatively weak
interactions, where the physics can be described well in terms of
mean-field or---in lattice geometries---simple lowest-band models.
However, interatomic forces can be experimentally tuned to a large
extent via Feshbach resonances \cite{koehler06}. In particular, in
quasi-one-dimensional systems, which emerge under strong transversal
confinement, it is possible to exploit confinement-induced resonances
\cite{Olshanii1998a} to explore the regime of strong correlations
\cite{kinoshita04,paredes04}. For infinitely repulsive bosons, this
is known as the fermionization limit, in allusion to the fact that
the system can be mapped exactly to an ideal Fermi gas \cite{girardeau60}.
Here the exclusion principle in a sense emulates the effect of \emph{hard-core}
interactions, to the extent that the bosons share local aspects with
their fermionic counterparts, whereas nonlocal properties such as
their coherence and momentum distribution are very different. The
basic crossover from the weakly interacting trapped Bose gas to the
fermionization limit had been predicted from a thermodynamic-limit
perspective \cite{petrov00,dunjko01} and interpreted in terms of
a mean-field picture \cite{alon05}. By contrast, it is only recently
that its \emph{microscopic} mechanism has been investigated within
an ab-initio framework \cite{hao06,zoellner06a,zoellner06b,deuretzbacher06}.

In this work, we tackle the obvious question of how the fermionization
crossover for the one-component Bose gas extends to a trapped \emph{two-component}
mixture. By way of analogy, tuning the \emph{inter-component} coupling
strength to the infinitely repulsive regime (for fixed intra-species
interactions) may be regarded as \emph{composite fermionization}.
Here, a recent study has extended the standard fermionization map
to mixtures of two \emph{identical} particle species with both intra-
and inter-species \emph{hard-core} interactions \cite{girardeau07}.
Apart from this special borderline case, little is known except for
a classification of the low-energy modes in the harmonic-fluid approximation
\cite{cazalilla03}. Here we study the crossover from weak to strongly
repulsive couplings between two components under harmonic confinement.
We will show that this composite fermionization can lead to demixing,
and lay out how it depends on the intra-species interactions, on the
densities of the two components, as well as on the masses and trapping
parameters of each species. 

Our paper is organized as follows. Section~\ref{sec:theory} introduces
the model and briefly reviews the fermionization map and its extension
to mixtures. In Sec.~\ref{sec:method}, we give a concise presentation
of the computational method. Section~\ref{sec:sym} first explores
the completely symmetric setup, where both components have equal atom
numbers, interaction constants, masses, and see the same harmonic
trap. The subsequent Sec.~\ref{sec:sym} in turn shows what different
phase-separation scenarios emerge if these constraints are relaxed
one by one.

\section{Theoretical background \label{sec:theory}}

\subsection{Model}

We consider a mixture of two distinguishable bosonic species, which
we shall label {}``A'' and {}``B''. These may correspond to atoms
with unequal nucleon numbers---be it different isotopes or altogether
different species---or possibly different hyperfine states of one
and the same species. Furthermore, we assume these to be confined
to quasi-one dimension (1D), such that the transverse degrees of freedom
may be integrated out. The effective low-energy Hamiltonian for an
arbitrary mixture of $N=N_{\mathrm{A}}+N_{\mathrm{B}}$ atoms then
reads \[
H=\sum_{\sigma=\mathrm{A,B}}H_{\sigma}+H_{\mathrm{AB}},\]
where the single-species Hamiltonian $H_{\sigma}$ and the inter-species
coupling $H_{\mathrm{AB}}$ read \begin{eqnarray*}
H_{\sigma} & = & \sum_{i=1}^{N_{\sigma}}\left[\frac{p_{\sigma,i}^{2}}{2M_{\sigma}}+U_{\sigma}(x_{\sigma,i})\right]+\sum_{i<j}g_{\sigma}\delta(x_{\sigma,i}-x_{\sigma,j})\\
H_{\mathrm{AB}} & = & \sum_{a=1}^{N_{\mathrm{A}}}\sum_{b=1}^{N_{\mathrm{B}}}g_{\mathrm{AB}}\delta(x_{\mathrm{A},a}-x_{\mathrm{B},b}).\end{eqnarray*}
Here we consider harmonic trapping potentials $U_{\sigma}(x)=\frac{1}{2}M_{\sigma}\omega_{\sigma}^{2}x^{2}$.
By rescaling to harmonic-oscillator units $a_{\mathrm{A}}\equiv\sqrt{\hbar/M_{\mathrm{A}}\omega_{\mathrm{A}}}$,
one can eliminate $M_{\mathrm{A}}=\omega_{\mathrm{A}}=1$ by exploiting
the scaling \[
H_{\mathrm{A}}(M_{\mathrm{A}},\omega_{\mathrm{A}},g_{\mathrm{A}};X_{\mathrm{A}})=\hbar\omega_{\mathrm{A}}H_{\mathrm{A}}(1,1,{g'}_{\mathrm{A}};{X'}_{\mathrm{A}}),\]
 with $X{}_{\mathrm{A}}'\equiv(x_{\mathrm{A},1}^{\prime},\dots,x_{\mathrm{A},N_{\mathrm{A}}}^{\prime})\equiv X_{\mathrm{A}}/a_{\mathrm{A}}$
and $g'_{\mathrm{A}}\equiv g_{\mathrm{A}}\sqrt{M_{\mathrm{A}}/\hbar^{3}\omega_{\mathrm{A}}}$
.

For numerical reasons, we regularize the delta-function interaction
by a normalized Gaussian of width much smaller than the inter-particle
distance; see Ref.~\cite{zoellner06a} for details.

\subsection{Fermionization}

The (single-component) 1D Bose gas has the peculiar property that
it is isomorphic to a system of identical fermions. In particular,
the \emph{standard} Bose-Fermi map relates the many-body wave function
of hard-core bosons (obeying the boundary condition \textbf{$\left.\Psi\right|_{x_{i}=x_{j}}=0,\, i<j$},
which corresponds to taking the 1D interaction strength $g\to\infty$)
to that of non-interacting fermions $\Psi_{-}$: \[
\Psi=A\Psi_{-},\qquad A(X):=\prod_{i<j}\mathrm{sgn}(x_{i}-x_{j}).\]
Specifically, the ground state is given simply by the absolute value
of the non-interacting fermionic ground state, $\Psi_{0}=\left|\Psi_{-,0}\right|$.
This makes it tempting to think of Pauli's exclusion principle as
emulating the effect of the repulsive interactions (or vice versa),
which is why the limit $g\to\infty$ is commonly referred to as \emph{fermionization}.
Note that, since $A^{2}=1$, all local quantities will coincide with
that computed from the fermion state. Specifically, this is the case
for the density $\rho_{N}=\left|\Psi\right|^{2}$ and any derived
quantities, such as the reduced (one- or two-body) densities. However,
nonlocal quantities such as the momentum distribution, may differ
dramatically from the fermionic ones.

The standard Bose-Fermi map above has recently been extended to mixtures
of two \emph{identical} species $\mathrm{A=B}$ (i.e., equal masses
and potentials) with hard-core \emph{intra-} and \emph{inter-species}
interactions, $g_{\sigma}=g_{\mathrm{AB}}\to\infty$. Its wave function
$\Psi$ is transformed to that of a system of $N=\sum_{\sigma}N_{\sigma}$
identical fermions \cite{girardeau07}. For the special case of A
and B being bosonic, this \emph{generalized} Bose-Fermi map $\Psi=A\Psi_{-}$
reads \begin{equation}
A(X_{\mathrm{A}},X_{\mathrm{B}})=A_{\mathrm{A}}(X_{\mathrm{A}})A_{\mathrm{B}}(X_{\mathrm{B}})A_{\mathrm{AB}}(X_{\mathrm{A}},X_{\mathrm{B}}),\label{eq:BFmap}\end{equation}
 where $A_{\sigma}(X_{\sigma})\equiv\prod_{1\le i<j\le N_{\sigma}}\mathrm{sgn}(x_{\sigma,i}-x_{\sigma,j})$
is the standard map restricted to subsystem $\sigma$, and $A_{\mathrm{AB}}(X_{\mathrm{A}},X_{\mathrm{B}})\equiv\prod_{a=1}^{N_{\mathrm{A}}}\prod_{b=1}^{N_{\mathrm{B}}}\mathrm{sgn}(x_{\mathrm{A},a}-x_{\mathrm{B},b})$
serves to impose hard-core boundary conditions on inter-species \emph{}collision
points. In the case of harmonic trapping, where the single-particle
orbitals are known analytically, the solution may even be written
down explicitly \cite{girardeau07}\begin{equation}
\Psi\left(X\equiv(X_{\mathrm{A}},X_{\mathrm{B}})\right)\propto e^{-|X|^{2}/2}\negthickspace\prod_{1\le i<j\le N}\negthickspace|x_{i}-x_{j}|.\label{eq:BF-HO}\end{equation}

\section{Computational method\label{sec:method}}

Our approach relies on the numerically exact Multi-Configuration Time-Dependent
Hartree method \cite{mey90:73,bec00:1}, a quantum-dynamics tool which
has been applied successfully to systems of few identical bosons (see
\cite{zoellner06a,zoellner06b,zoellner07a,zoellner07c,zoellner07b}).
Its principal idea is to solve the time-dependent Schrödinger equation
$\begin{array}{c}
i\dot{\Psi}(t)=H\Psi(t)\end{array}$ as an initial-value problem by expanding the solution in terms of
direct (or Hartree) products $\Phi_{J}\equiv\varphi_{j_{1}}^{(1)}\otimes\cdots\otimes\varphi_{j_{N}}^{(N)}$:\begin{equation}
\Psi(t)=\sum_{J}A_{J}(t)\Phi_{J}(t).\label{eq:mctdh-ansatz}\end{equation}
The (unknown) \emph{single-particle functions} $\varphi_{j}^{(\kappa)}$
($j=1,\dots,n_{\kappa}$) are in turn represented in a fixed \emph{primitive}
basis implemented on a grid. In our case of, where particles of each
species are indistinguishable, the single-particle functions within
each subset $\kappa\in\{1,\dots,N_{\mathrm{A}}\}$ and $\{ N_{\mathrm{A}}+1,\dots,N\}$
are of course identical (i.e., we have $\{\varphi_{j_{\sigma}}^{(\sigma)}\}$,
with $j_{\sigma}\le n_{\sigma}$). This, along with the correct symmetrization
of the expansion coefficients $A_{J}$, ensures permutation symmetry
within each subset A,B. 

Note that in the above expansion not only the coefficients $A_{J}$
but also the single particle functions $\varphi_{j}$ are time dependent.
Using the Dirac-Frenkel variational principle, one can derive equations
of motion for both $A_{J},\varphi_{j}$ \cite{bec00:1}. Integrating
this differential-equation system allows us to obtain the time evolution
of the system via (\ref{eq:mctdh-ansatz}). This has the advantage
that the basis $\{\Phi_{J}(t)\}$ is variationally optimal at each
time $t$. Thus it can be kept relatively small, rendering the procedure
very efficient. 

Although designed for time-dependent simulations, it is also possible
to apply this approach to stationary states. This is done via the
so-called \emph{relaxation method} \cite{kos86:223}. The key idea
is to propagate some wave function $\Psi(0)$ by the non-unitary $e^{-H\tau}$
(\emph{propagation in imaginary time}.) As $\tau\to\infty$, this
exponentially damps out any contribution but that stemming from the
true ground state like $e^{-(E_{m}-E_{0})\tau}$. In practice, one
relies on a more sophisticated scheme termed \emph{improved relaxation}
\cite{mey03:251}, which is much more robust especially for excitations.
Here $\langle\Psi|H|\Psi\rangle$ is minimized with respect to both
the coefficients $A_{J}$ and the orbitals $\varphi_{j}$. The effective
eigenvalue problems thus obtained are then solved iteratively by first
solving for $A_{J}$ with \emph{fixed} orbitals and then {}`optimizing'
$\varphi_{j}$ by propagating them in imaginary time over a short
period. That cycle will then be repeated.

\section{Composite-fermionization transition \label{sec:sym}}

In contrast to the case of a single bosonic species, binary mixtures
offer a plethora of different parameters, making the physics richer
and less straightforward: In principle, we may have different atom
numbers $N=N_{\mathrm{A}}+N_{\mathrm{B}}$, different masses $M_{\sigma=\mathrm{A,B}}$,
intra- and inter-species couplings $g_{\sigma}$ ($g_{\mathrm{AB}}$),
and species-dependent traps $U_{\sigma}(x)$. In this section, in
order to illustrate the basic mechanism of the crossover from weak
to strongly repulsive inter-species interactions, $g_{\mathrm{AB}}\in[0,\infty)$,
we focus on the simplest, symmetric setup where\[
N_{\sigma}=\frac{N}{2};\; M_{\sigma}=1;\; g_{\sigma}=g;\; U_{\sigma}(x)=\frac{1}{2}x^{2}\quad(\sigma=\mathrm{A,B}).\]
In this case, $H$ has an exact permutation symmetry between species
A and B. This idealized situation may correspond to two internal states
of the same species or, ignoring slight mass deviations, two different
isotopes, where $g_{\mathrm{AB}}$ is tuned via the inter-species
scattering length. Actually, in the special case where $g_{\sigma}=g_{\mathrm{AB}}=g$,
this system maps to a one-component Bose gas with $N=\sum_{\sigma}N_{\sigma}$
atoms \cite{zoellner06a,zoellner06b,deuretzbacher06} (for any number
of components and any $N_{\sigma}$, for that matter) -- up to permutational
degeneracies, which are not that severe for the ground state.

\begin{figure}
\begin{centering}\includegraphics[width=0.8\columnwidth,keepaspectratio]{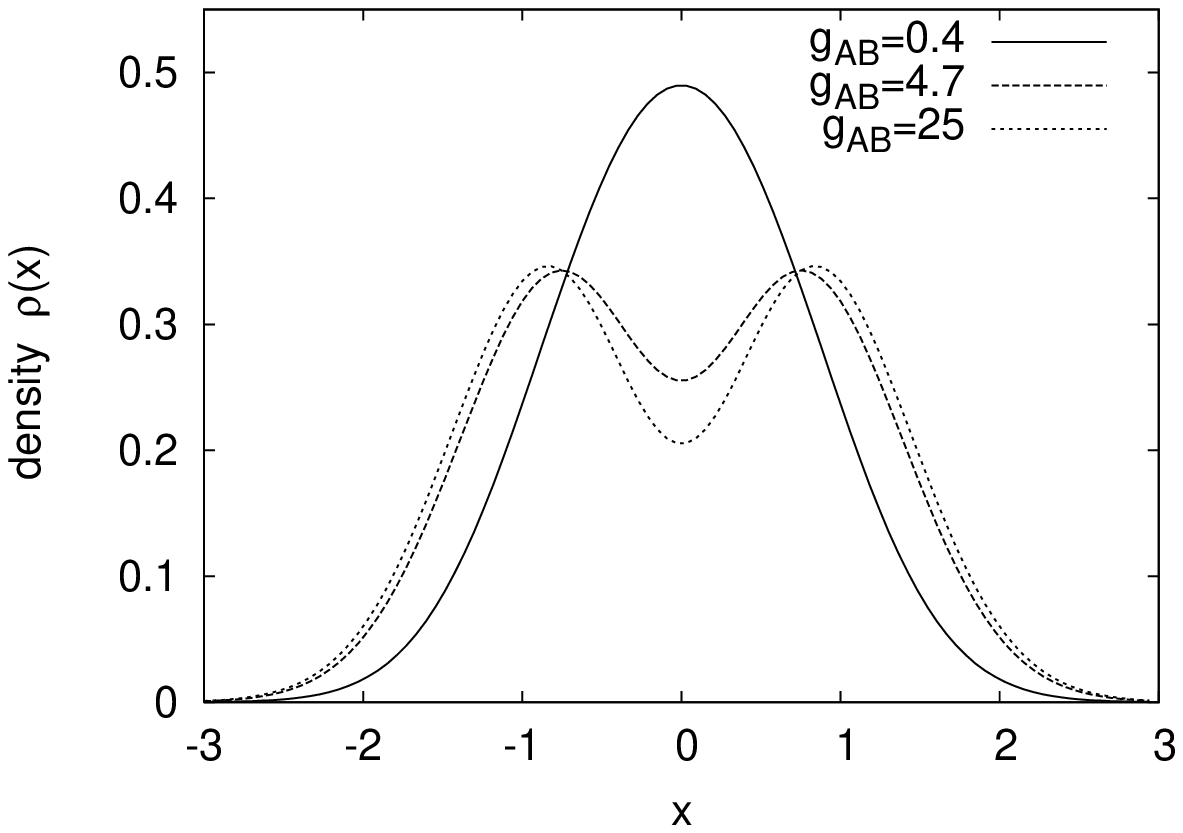}\par\end{centering}

\begin{centering}\includegraphics[width=0.33\columnwidth,keepaspectratio]{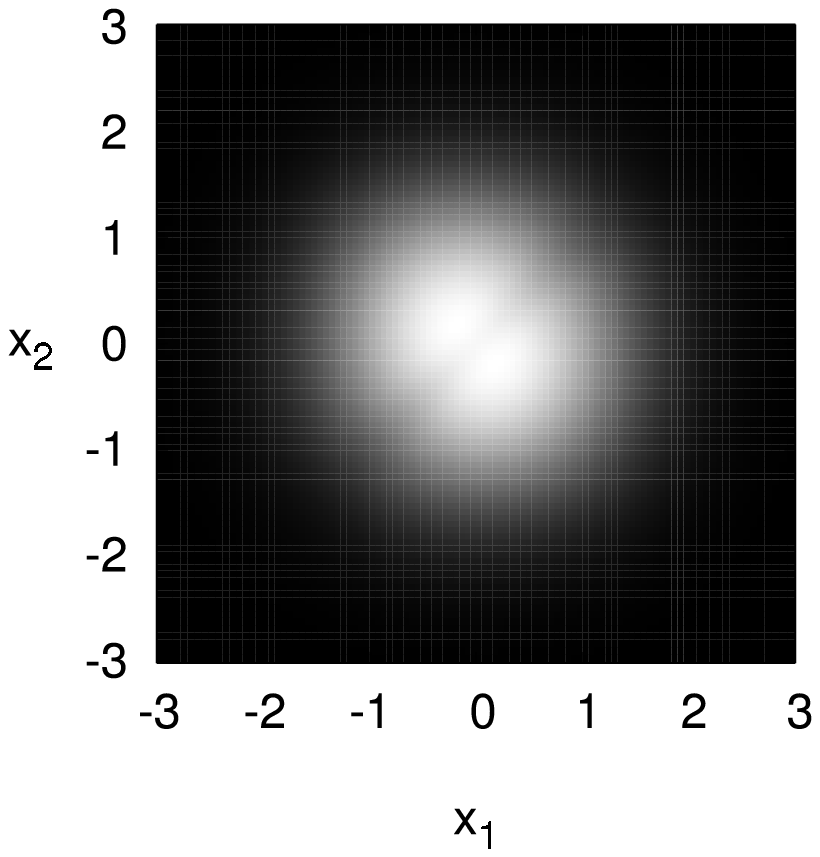}\includegraphics[width=0.33\columnwidth,keepaspectratio]{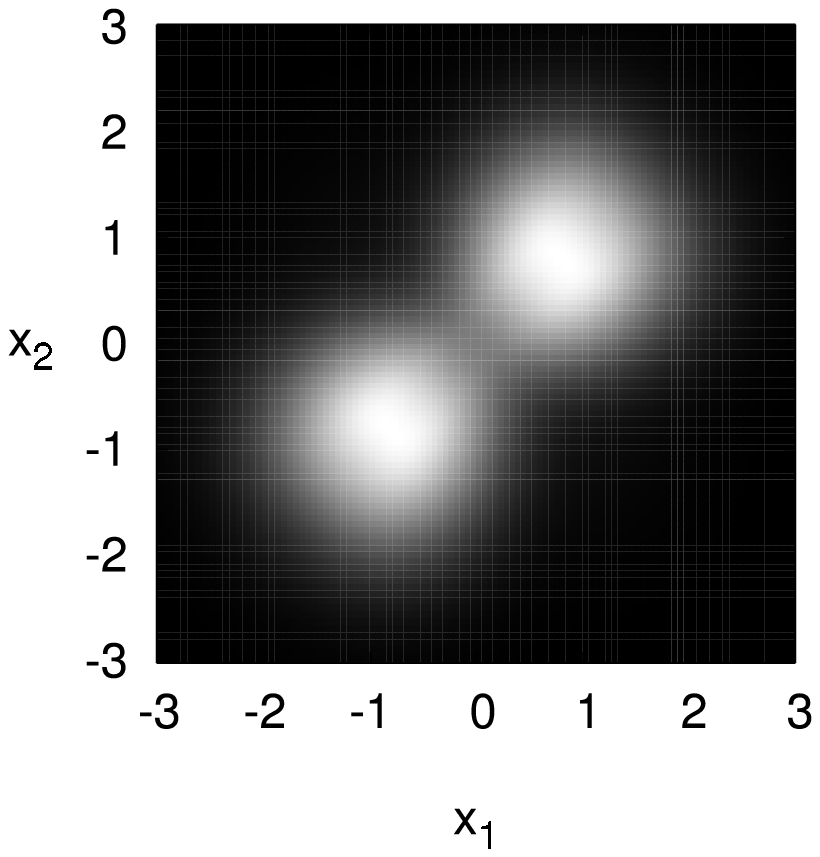}\includegraphics[width=0.33\columnwidth,keepaspectratio]{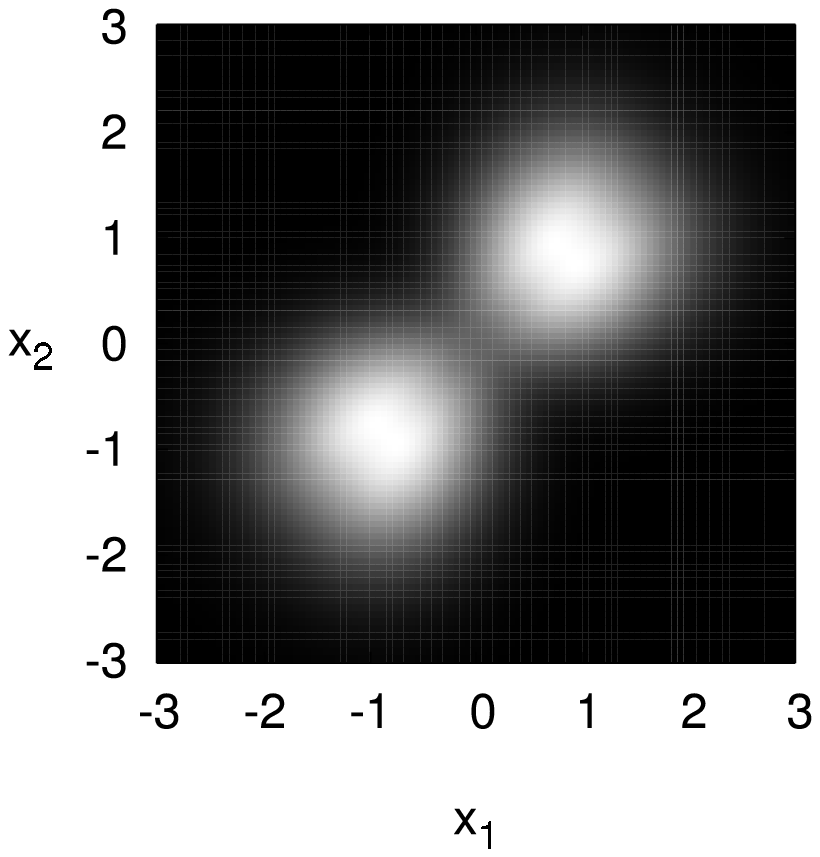}\par\end{centering}

\begin{centering}\includegraphics[width=0.33\columnwidth,keepaspectratio]{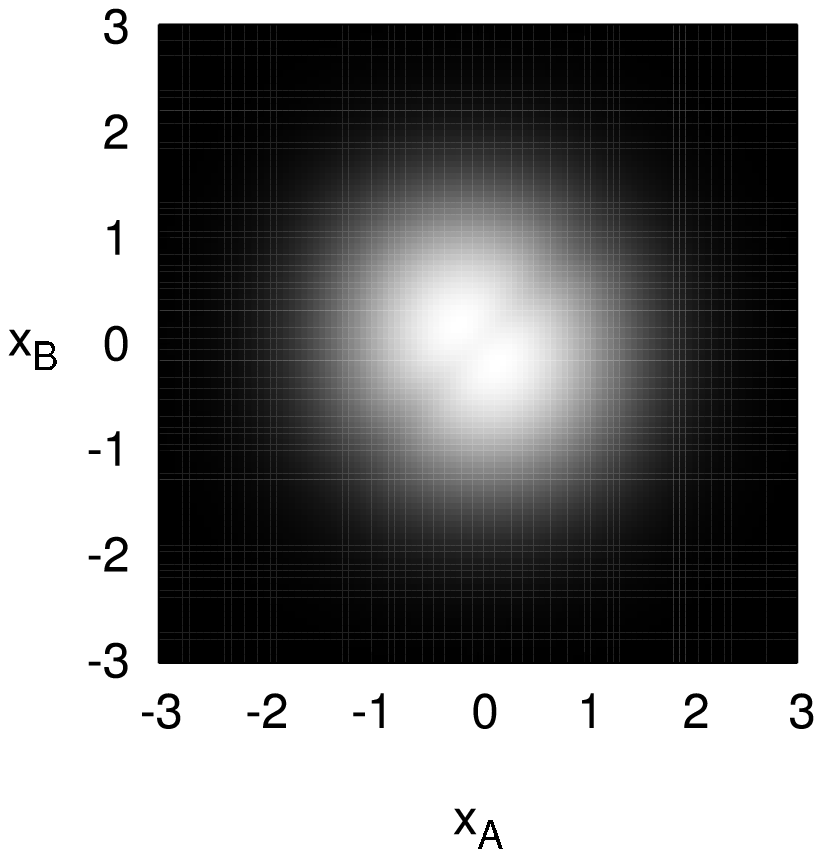}\includegraphics[width=0.33\columnwidth,keepaspectratio]{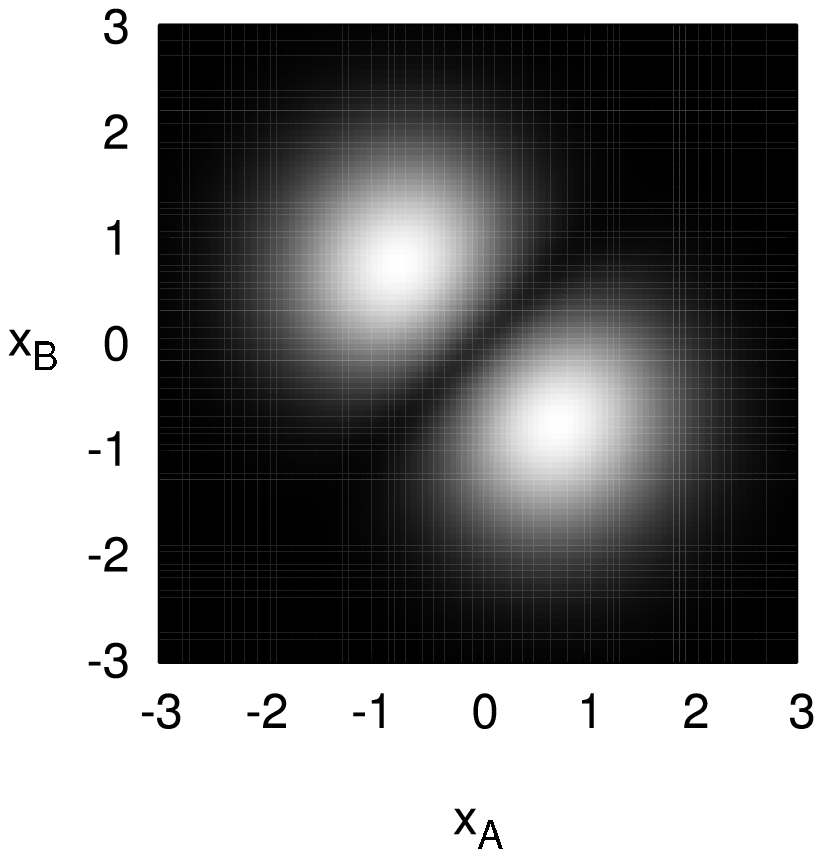}\includegraphics[width=0.33\columnwidth,keepaspectratio]{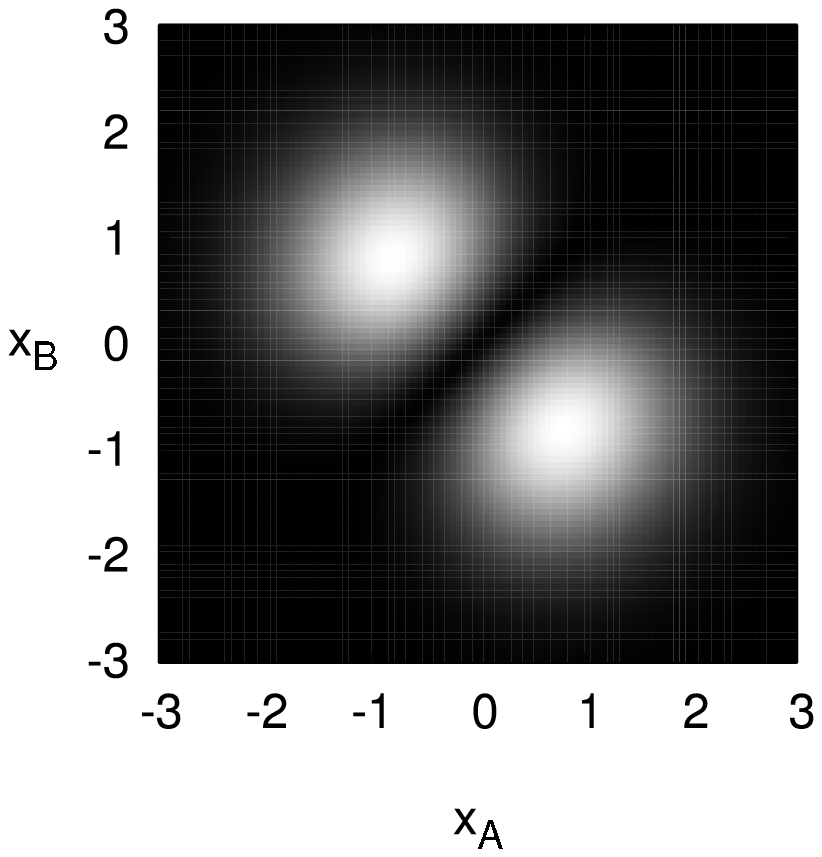}\par\end{centering}

\caption{Composite fermionization of a mixture with \emph{$N_{\mathrm{\sigma=A,B}}=2$}
bosons with intra-component interaction $g_{\sigma}=0.4$. \emph{Top}:
density profiles $\rho_{\mathrm{}}(x)$; \emph{Bottom}: Two-body correlation
functions $\rho_{\mathrm{\sigma\sigma}}(x_{1},x_{2})$ and $\rho_{\mathrm{AB}}(x_{\mathrm{A}},x_{\mathrm{B}})$
for inter-species couplings $g_{\mathrm{AB}}=0.4,\,4.7$ and $25$
(from left to right). \label{cap:2+2-g0.4}}
\end{figure}
\begin{figure}
\begin{centering}\includegraphics[width=0.8\columnwidth,keepaspectratio]{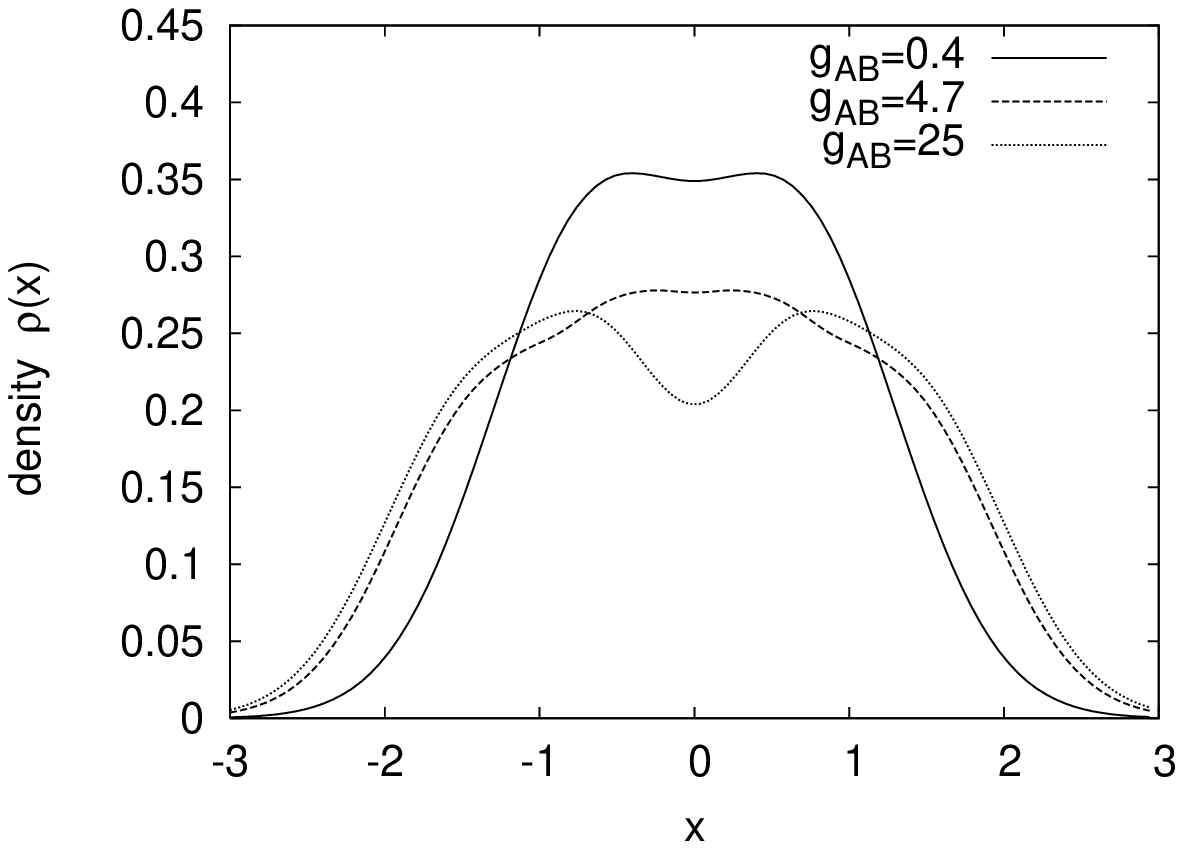}\par\end{centering}

\includegraphics[width=0.33\columnwidth,keepaspectratio]{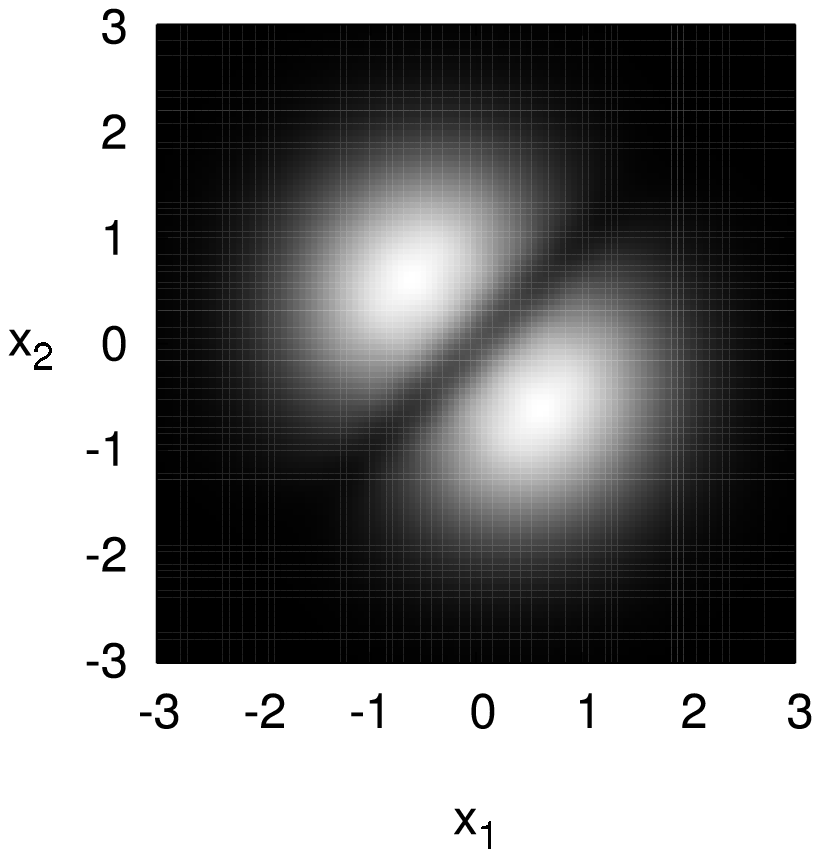}\includegraphics[width=0.33\columnwidth,keepaspectratio]{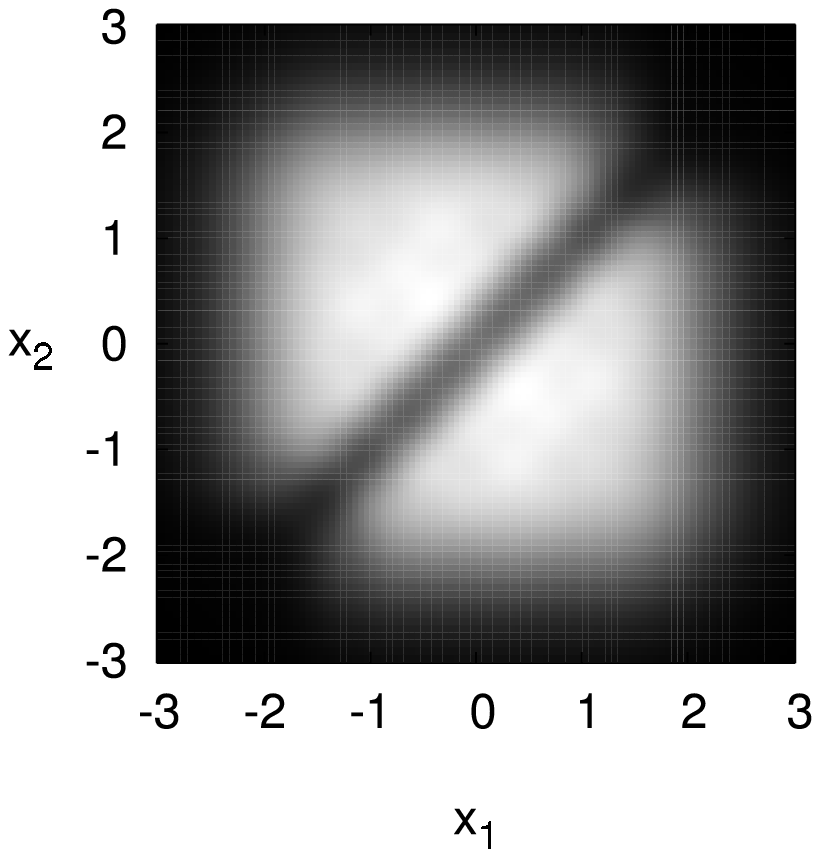}\includegraphics[width=0.33\columnwidth,keepaspectratio]{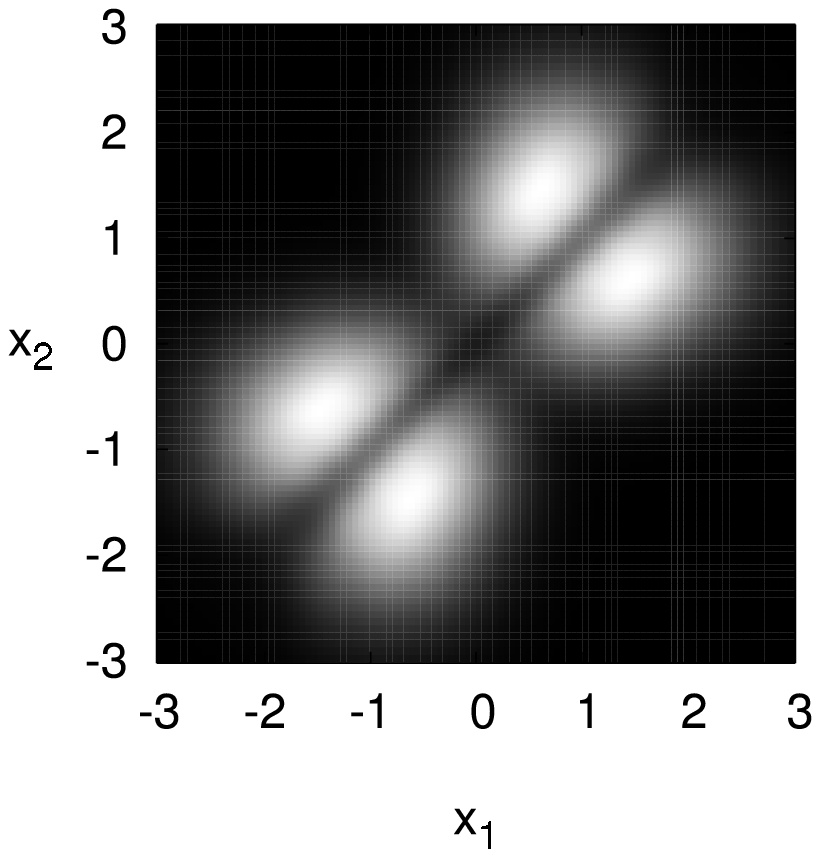}

\includegraphics[width=0.33\columnwidth,keepaspectratio]{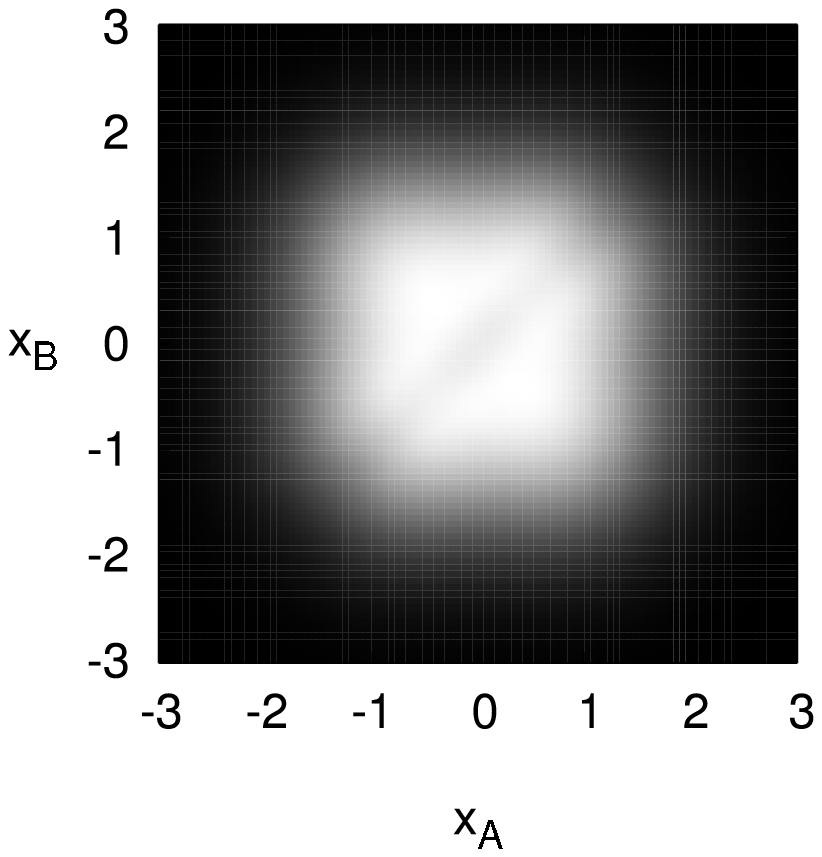}\includegraphics[width=0.33\columnwidth,keepaspectratio]{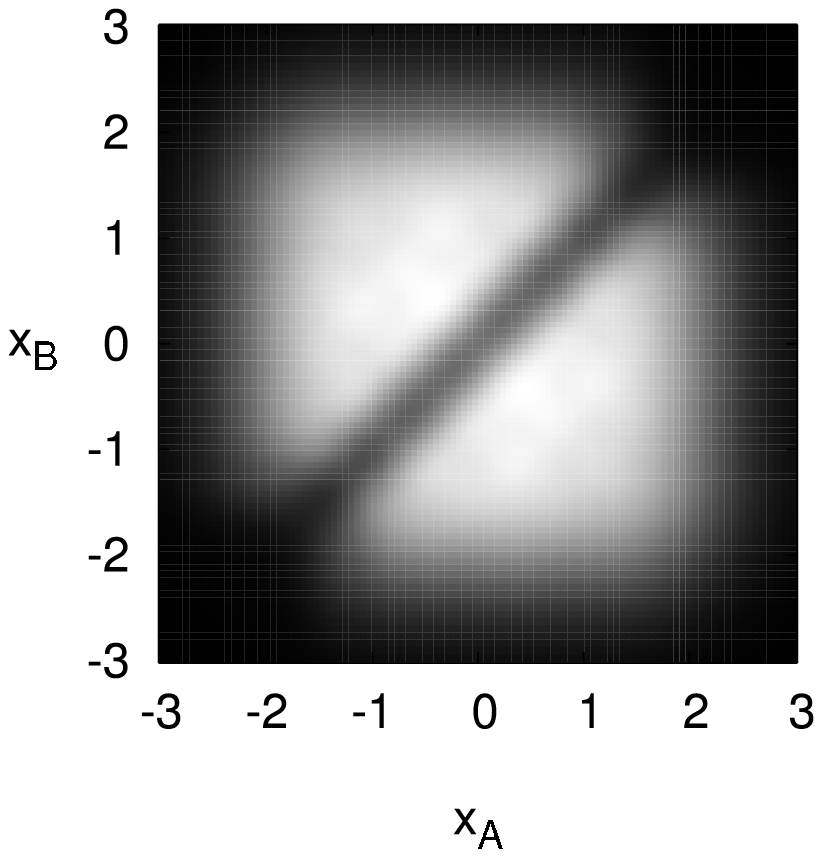}\includegraphics[width=0.33\columnwidth,keepaspectratio]{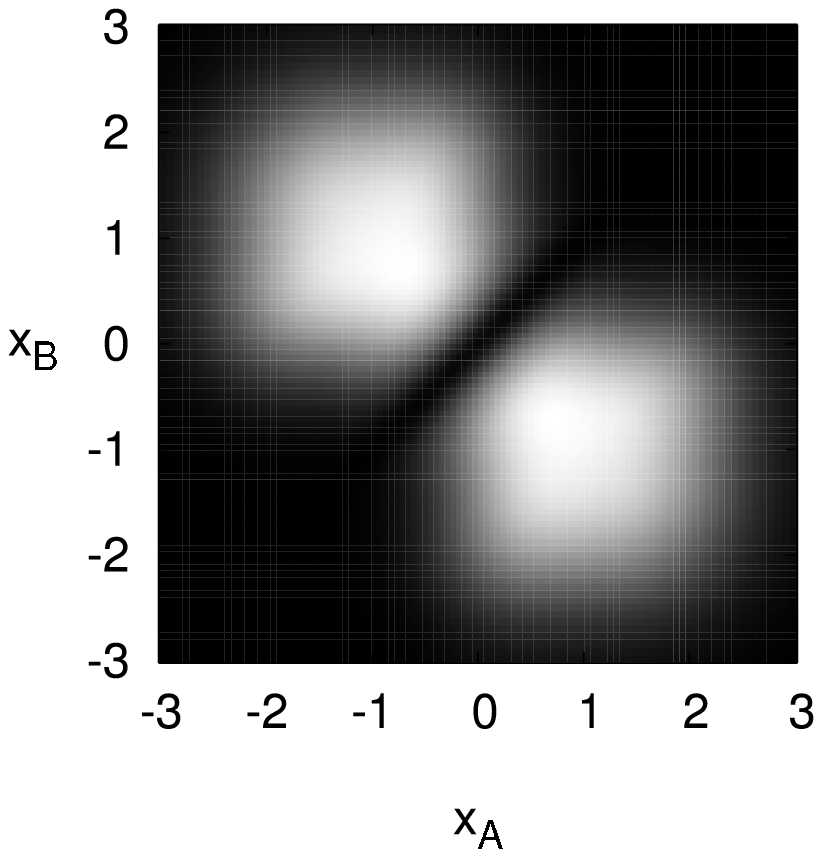}

\caption{Same as Fig.~\ref{cap:2+2-g0.4}, but with $g_{\mathrm{A}}=g_{\mathrm{B}}=4.7$.
\label{cap:2+2-g4.7}}
\end{figure}

\begin{figure}
\begin{centering}\includegraphics[width=0.8\columnwidth,keepaspectratio]{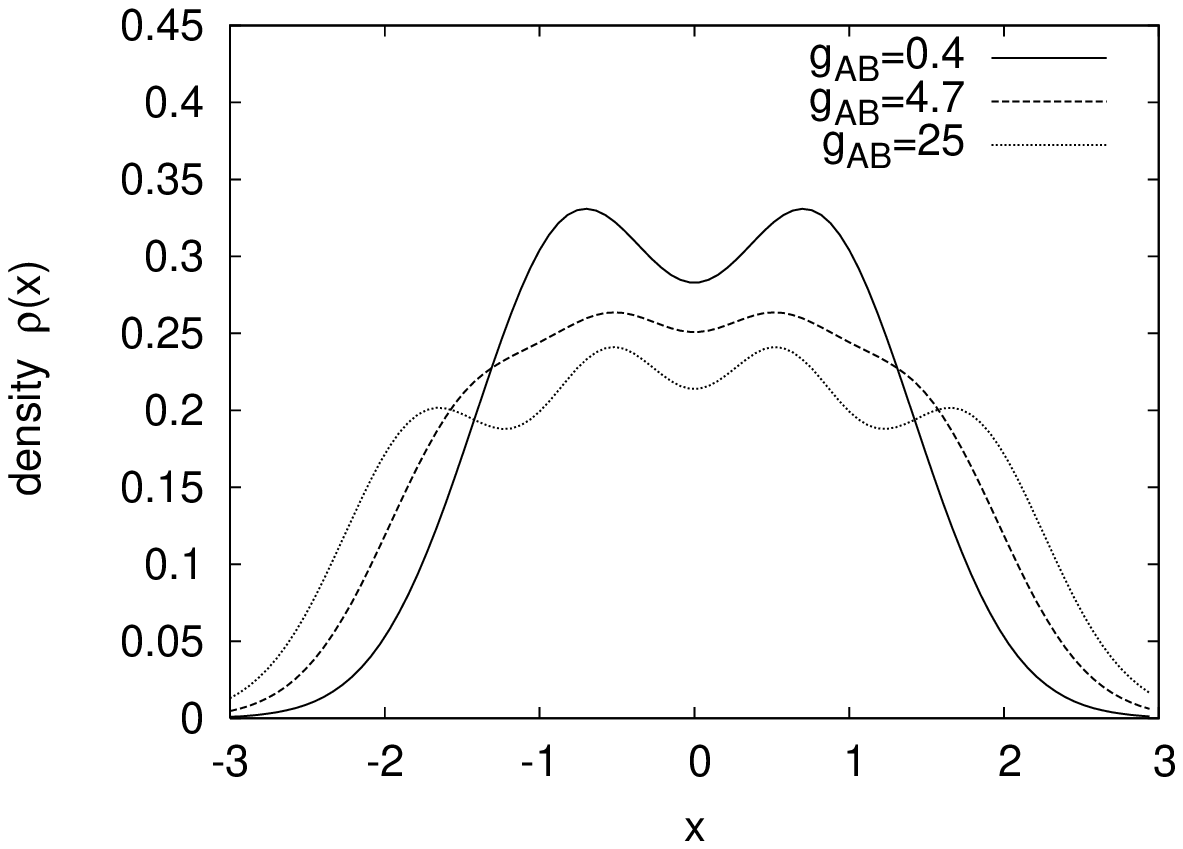}\par\end{centering}

\includegraphics[width=0.33\columnwidth,keepaspectratio]{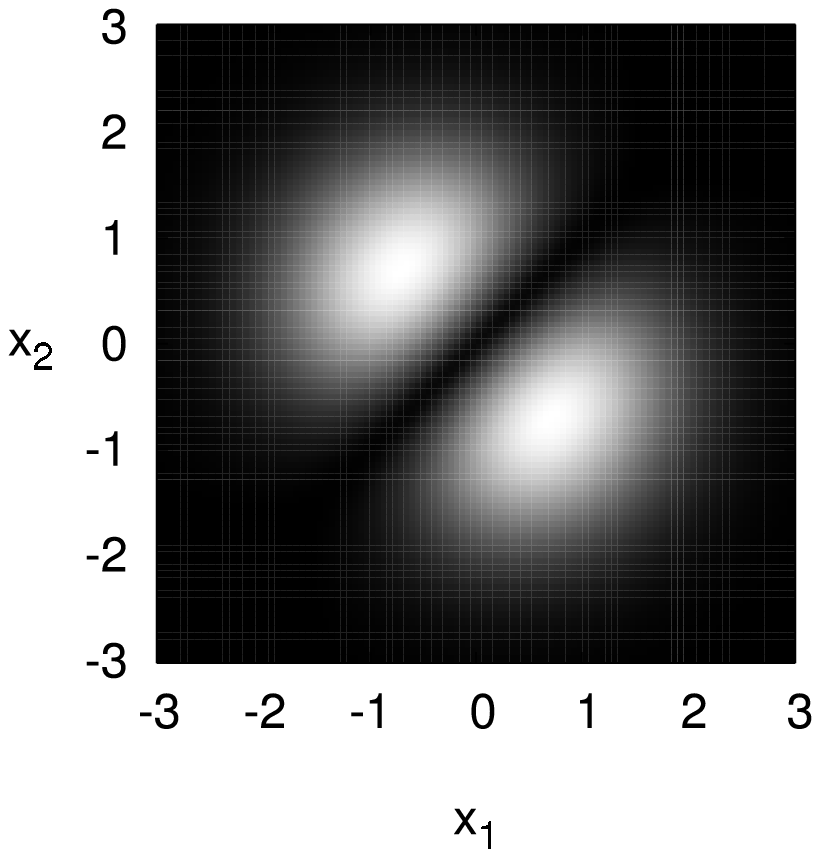}\includegraphics[width=0.33\columnwidth,keepaspectratio]{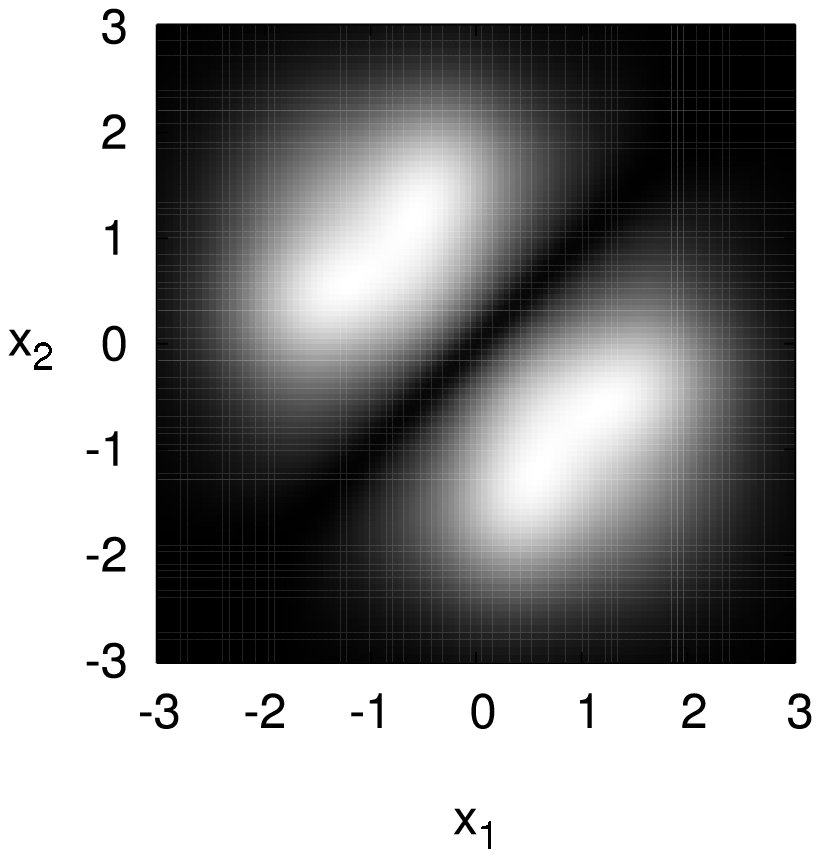}\includegraphics[width=0.33\columnwidth,keepaspectratio]{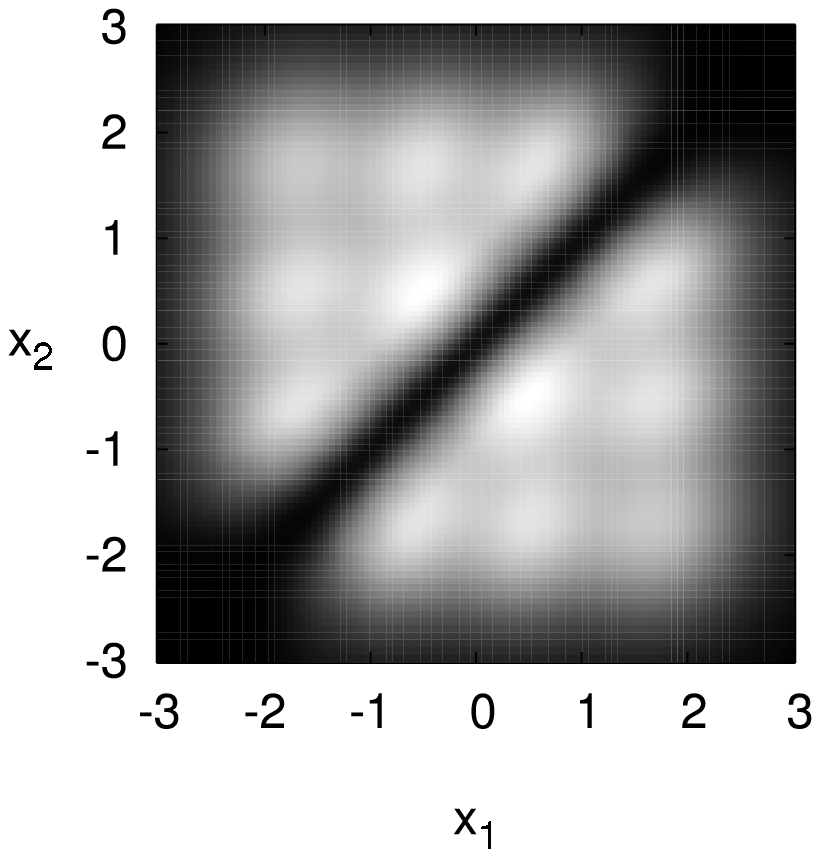}

\includegraphics[width=0.33\columnwidth,keepaspectratio]{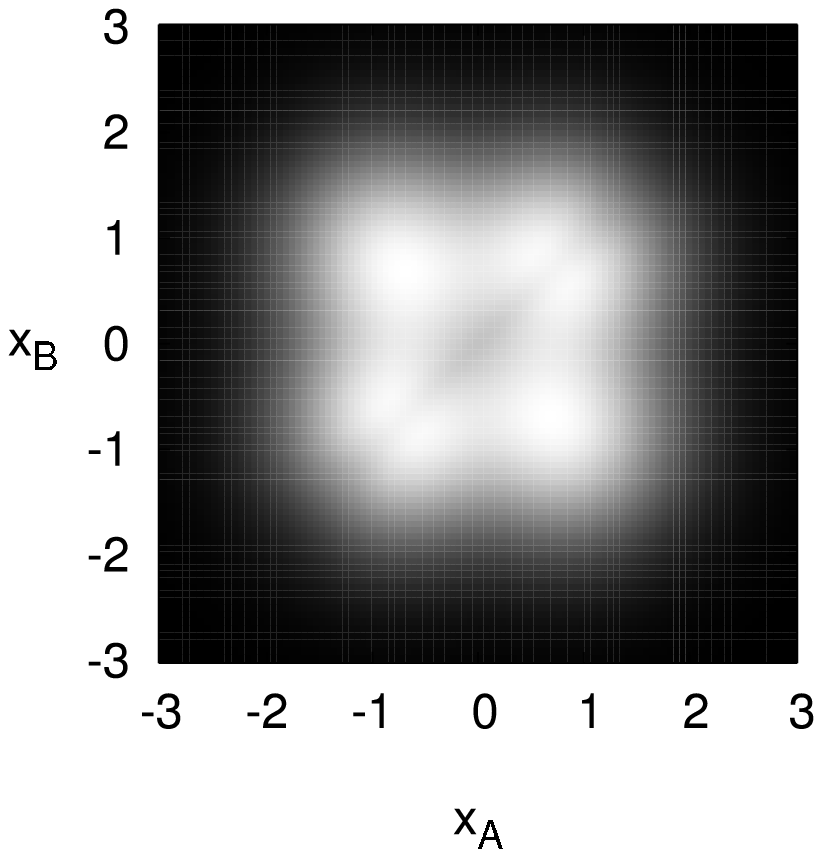}\includegraphics[width=0.33\columnwidth,keepaspectratio]{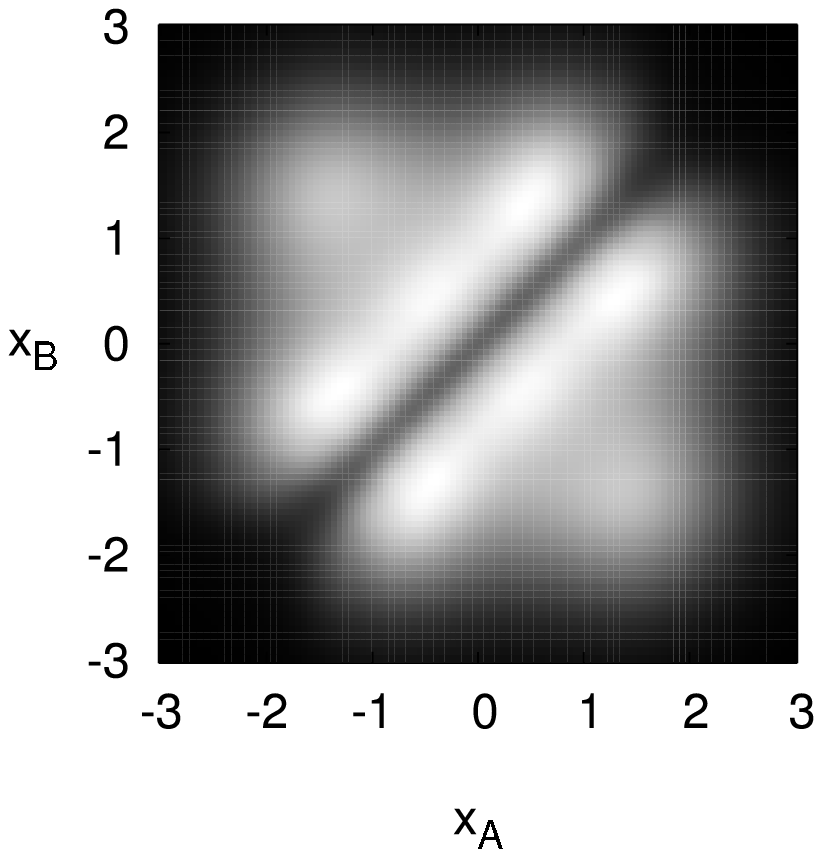}\includegraphics[width=0.33\columnwidth,keepaspectratio]{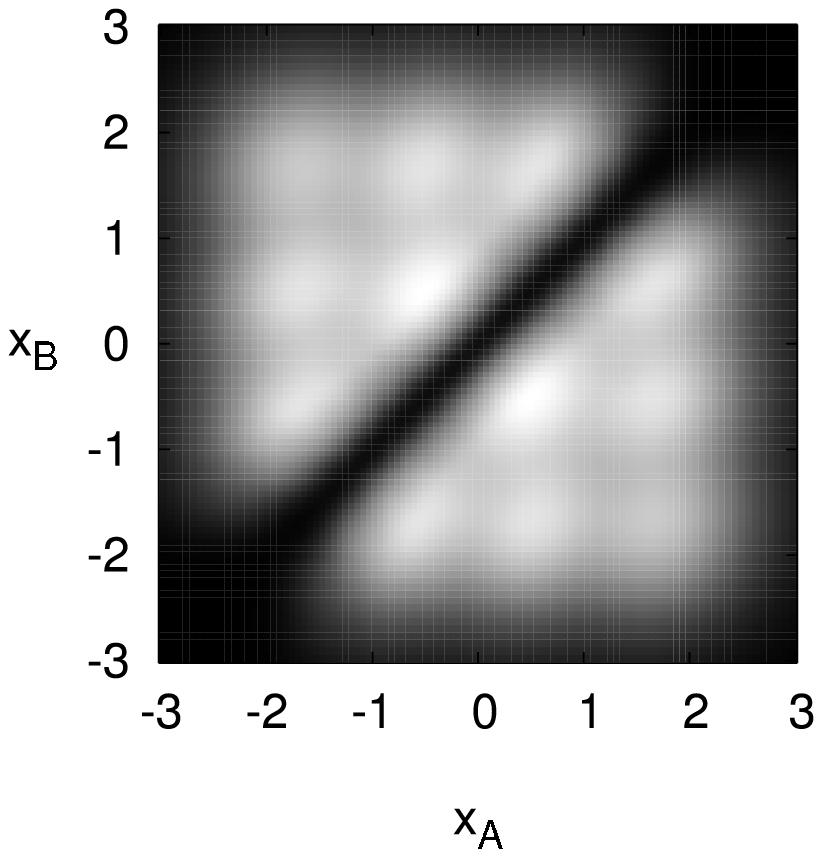}

\caption{Same as Fig.~\ref{cap:2+2-g0.4}, but with $g_{\mathrm{A}}=g_{\mathrm{B}}=25$.
\label{cap:2+2-g25}}
\end{figure}

Here, by contrast, we are interest in the following question: What
happens to the mixture--- $g_{\sigma}$ being fixed---when $g_{\mathrm{AB}}$
is varied up to the hard-core limit? (This we refer to as \emph{composite
fermionization} of the subsystems A and B, despite the general lack
of a Bose-Fermi mapping as Eq.~\ref{eq:BFmap}.) To get an impression
of that crossover, let us start with the case of two almost ideal
Bose gases, $g_{\sigma}=0.4$, each consisting of $N_{\sigma}=2$
atoms (similar results hold for larger atom numbers). Figure~\ref{cap:2+2-g0.4}
displays the evolution of the density profile $\rho(x)\equiv\rho_{\sigma}(x)$,
measuring the probability distribution for finding one $\sigma$ atom
at position $x$. Obviously, for $g_{\mathrm{AB}}\to0$, the total
state $\Psi=\Psi_{\mathrm{A}}\otimes\Psi_{\mathrm{B}}$ simply consists
of two uncorrelated {}``condensates'' ($\Psi_{\sigma}=\phi_{0}^{\otimes N_{\sigma}}$
for $g=0$), slightly smeared out due to repulsion. Increasing $g_{\mathrm{AB}}$
leads to an ever deeper dip in the profiles. This should be contrasted
with the case of two \emph{single} fermionized bosons, $N_{\sigma}=1$
\cite{cirone01}. The dip in Fig.~\ref{cap:2+2-g0.4} is much more
pronounced, which is indicative of phase separation, if symmetry screened:
$\rho_{\mathrm{A}}=\rho_{\mathrm{B}}$ are completely identical by
symmetry. However, this only corresponds to an ensemble average --
in a \emph{single} measurement, we will always find all $N_{\mathrm{A}}$
atoms on one side of the trap and $N_{\mathrm{B}}$ atoms on the other.
This claim is underscored by Fig.~\ref{cap:2+2-g0.4}(\emph{bottom}),
which reveals the evolution of the two-body densities $\rho_{\sigma,\sigma'}$.
If we were to measure, say, the first A-type boson at $x_{\mathrm{A},1}\approx1$,
then we are sure to find the second A boson also in that region $x_{\mathrm{A},2}\approx1$
and not on the left, and \emph{vice versa}. By contrast, the probability
for subsequently finding a B particle at the same position is virtually
zero, as dictated by the hard-core boundary condition, $\Psi|_{x_{\mathrm{A,}a}=x_{\mathrm{B},b}}=0$
($g_{\mathrm{AB}}\to\infty$). This makes it tempting to think of
this an entangled state of the form \[
|N_{\mathrm{A}},0\rangle\otimes|0,N_{\mathrm{B}}\rangle+|0,N_{\mathrm{A}}\rangle\otimes|N_{\mathrm{B}},0\rangle,\]
where $|n_{\mathrm{L}},n_{\mathrm{R}}\rangle\in\mathbb{H}_{\sigma}$
denotes a state with $n_{\mathrm{L,R}}$ atoms localized on the left
(right). It should be noted that, even for $g_{\sigma}=0$, there
is no simple mapping to fermions as (\ref{eq:BFmap}), since the hard-core
condition is imposed only on inter-species collision points, and thus
the information about which fragment the individual coordinates belong
to needs to be retained. However, in our special case of a harmonic
trap, it is natural to conjecture that the exact solution is given
by a modification of the single-species fermionization limit \cite{girardeau01},
\[
\Psi_{g_{\mathrm{AB}}\to\infty}(X)\propto e^{-|X|^{2}/2}\negthickspace\prod_{a\le N_{\mathrm{A}},b\le N_{\mathrm{B}}}\negthickspace|x_{\mathrm{A,}a}-x_{\mathrm{B},b}|,\]
which obeys the correct boundary conditions at points of inter-species
collisions. Trusting that logic, an analogous extension should hold
for the homogeneous system \cite{girardeau60}.

A similar pathway is encountered for two more strongly interacting
components, $g_{\sigma}=4.7$ (see Fig.~\ref{cap:2+2-g4.7}, \emph{top}).
At $g_{\mathrm{AB}}=0.4$, we have more or less two uncorrelated clouds,
which are governed by the desire to reduce their \emph{intra}-species
interaction energy. As $g_{\mathrm{AB}}=4.7$ reaches $g_{\sigma}$,
this turns into a trade-off between avoiding the own species just
as much as the other component. Letting $g_{\mathrm{AB}}\to\infty$,
the inter-particle repulsion takes over, and a similar phase-separation
tendency of A and B as before may be recognized in Fig.~\ref{cap:2+2-g4.7}.
In contrast to the {}``condensate'' case, however, the separation
of the two peaks is not pronounced as each hump is quite smeared out
in itself due to the intra-species repulsion. This is illuminated
further by the two-body densities (Fig.~\ref{cap:2+2-g4.7}): Here
the pattern for $\rho_{\sigma\sigma}(x_{1},x_{2})$ at $g_{\mathrm{AB}}=25$
is modulated by a correlation hole at $x_{1}=x_{2}$ due to intra-species
repulsion, as compared to the weakly interacting components (Fig.~\ref{cap:2+2-g0.4}).
This explains the two broadened peaks in $\rho_{\sigma}(x)$. 

So far, we have seen that the components tend to separate when the
inter-species repulsion overwhelms the intra-species one. This naturally
brings up the question of the fate of two initially fermionized components,
as shown for $g_{\sigma}=25$ in Fig.~\ref{cap:2+2-g25}. Notably,
by the conventional Bose-Fermi map, this relates to a Fermi-Fermi
mixture. Weak couplings $g_{\mathrm{AB}}=0.4$ pass the two fermionized
clouds largely unnoticed, which exhibit $N_{\sigma}=2$ characteristic
humps in $\rho_{\sigma}(x)$ \cite{girardeau01}. However, for larger
values $g_{\mathrm{AB}}=4.7$, the profiles slowly rearrange to a
more complex structure, which culminates in a profile with $N=4$
wiggles at $g_{\mathrm{AB}}=g_{\sigma}=25$. The density oscillations
signify that \emph{each} of the four atoms seeks an isolated spot,
irrespective of its species. That interpretation is supported by the
plots of the two-body densities $\rho_{\sigma\sigma}=\rho_{\mathrm{AB}}$,
which for $g_{\mathrm{AB}}=25$ reveal the checkerboard pattern familiar
from the single-boson crossover \cite{zoellner06a}. This should be
contrasted with the intermediate regime where $g_{\mathrm{AB}}=4.7<g_{\sigma}$:
Here two, say, A atoms are still localized on the left and on the
right side as for $g_{\mathrm{AB}}=0$. Upon measuring an A atom at,
say, $x_{\mathrm{A}}\approx1.5$, the two B atoms will likely be found
at either $x_{\mathrm{B}}\approx0.5$ or $x_{\mathrm{B}}\approx-1.5$,
this way remaining isolated from each other but also avoiding the
A atom. 

Note that, in agreement with our earlier remarks, the case $g_{\sigma}=g_{\mathrm{AB}}$
relates to a single-component Bose gas, which in turn maps to an ideal
\emph{Fermi} gas via (\ref{eq:BFmap}) in the limit $g_{\mathrm{AB}}\to\infty$.
As in that case, for $N\gg1$ these $N$ peaks become ever tinier
modulations on the envelope density, which for a harmonic trap can
be computed as $\bar{\rho}(x)=\sqrt{2N-x^{2}}/N\pi$ \cite{kolomeisky00}.

At this stage, we should point out that this limit is highly degenerate:
For one thing, there is a permutation degeneracy between A and B particles.
Second, in the limit $g_{\mathrm{AB}}\to\infty$, the ground-state
wave function (which is non-negative) degenerates with the fermionic
one by the Bose-Fermi map and, since no specific permutation symmetry
is imposed when treating the two components as distinguishable, all
solutions even with mixed A-B-exchange statistics are permissible
\cite{girardeau07}.

\paragraph*{Symmetry-breaking instability. \label{sub:Symmetry-break}}

\begin{figure}
\includegraphics[width=0.5\columnwidth,keepaspectratio]{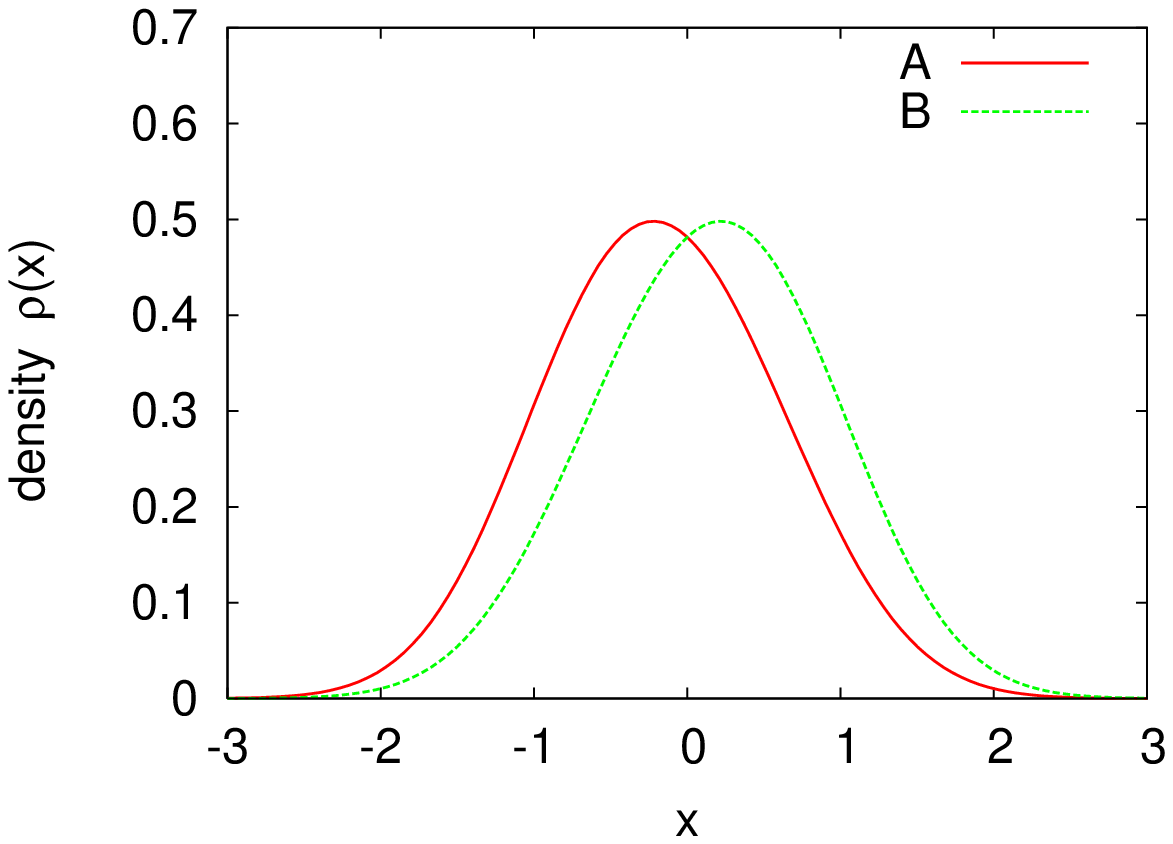}\includegraphics[width=0.5\columnwidth,keepaspectratio]{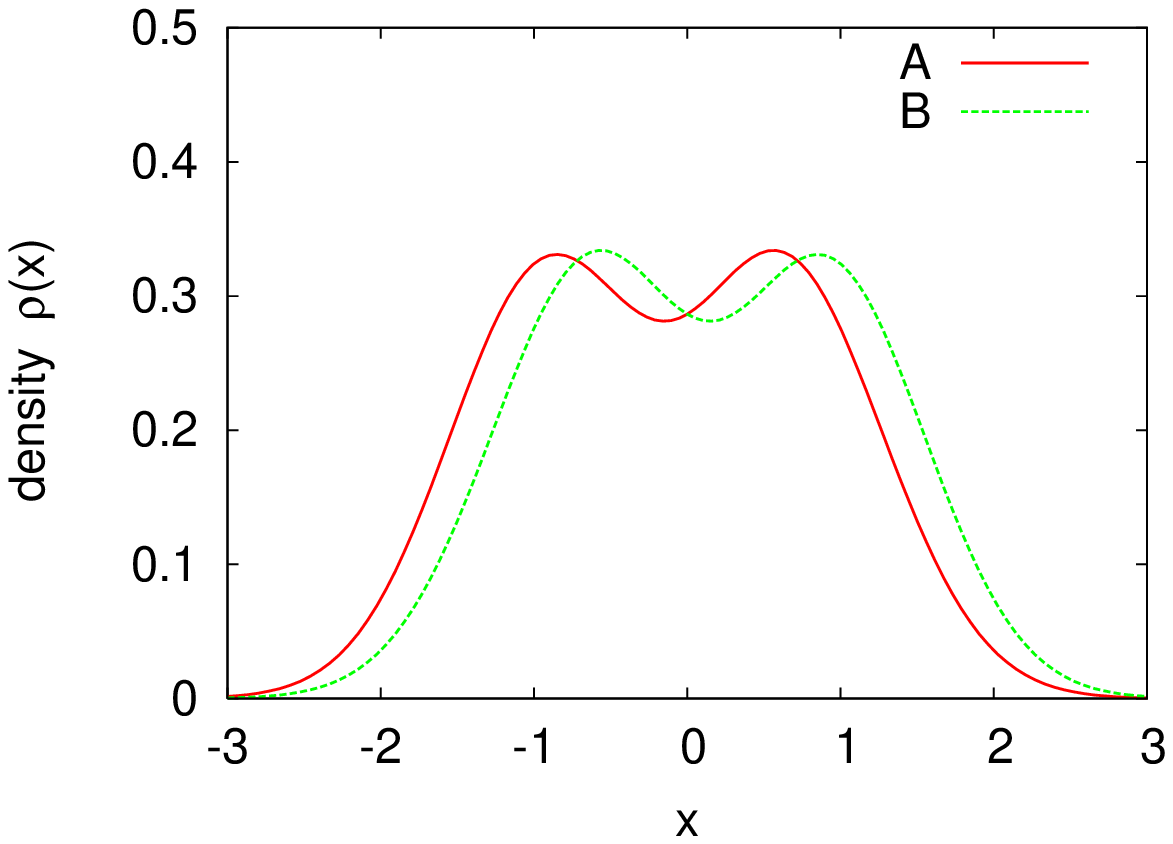}

\includegraphics[width=0.5\columnwidth,keepaspectratio]{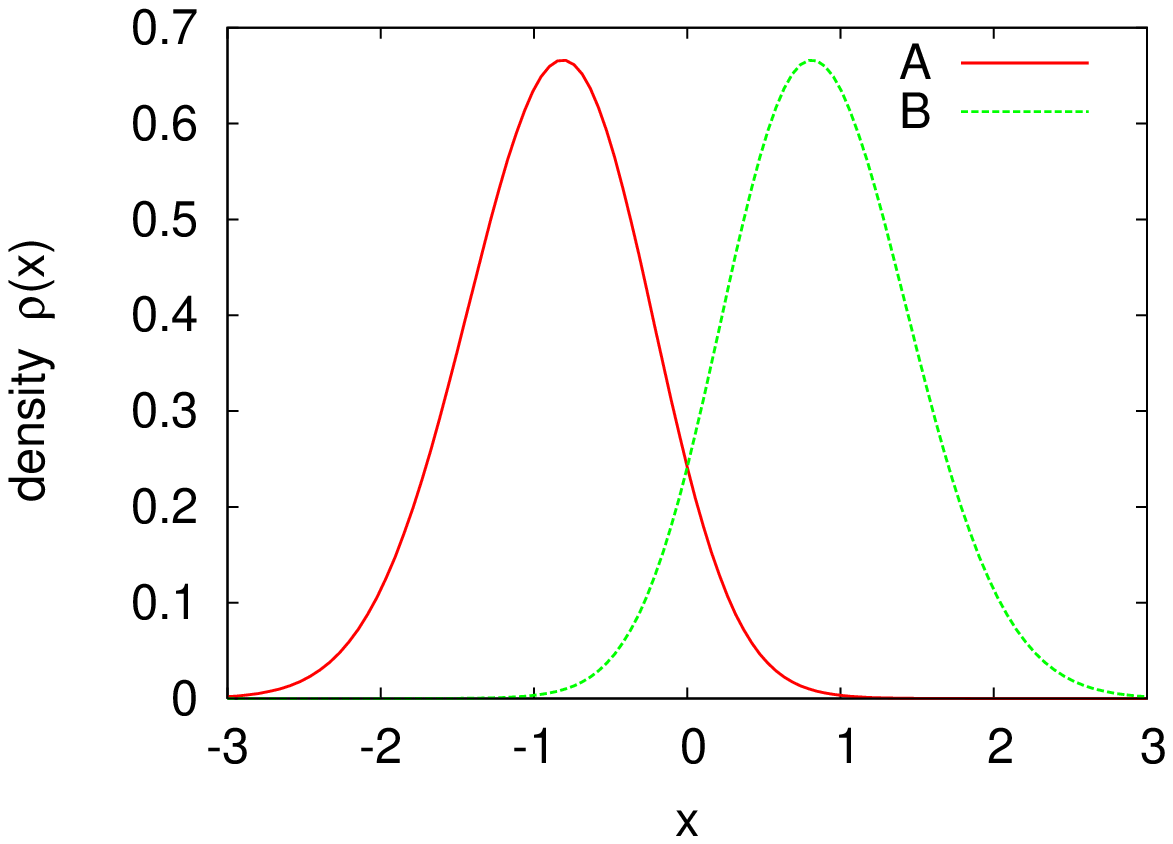}\includegraphics[width=0.5\columnwidth,keepaspectratio]{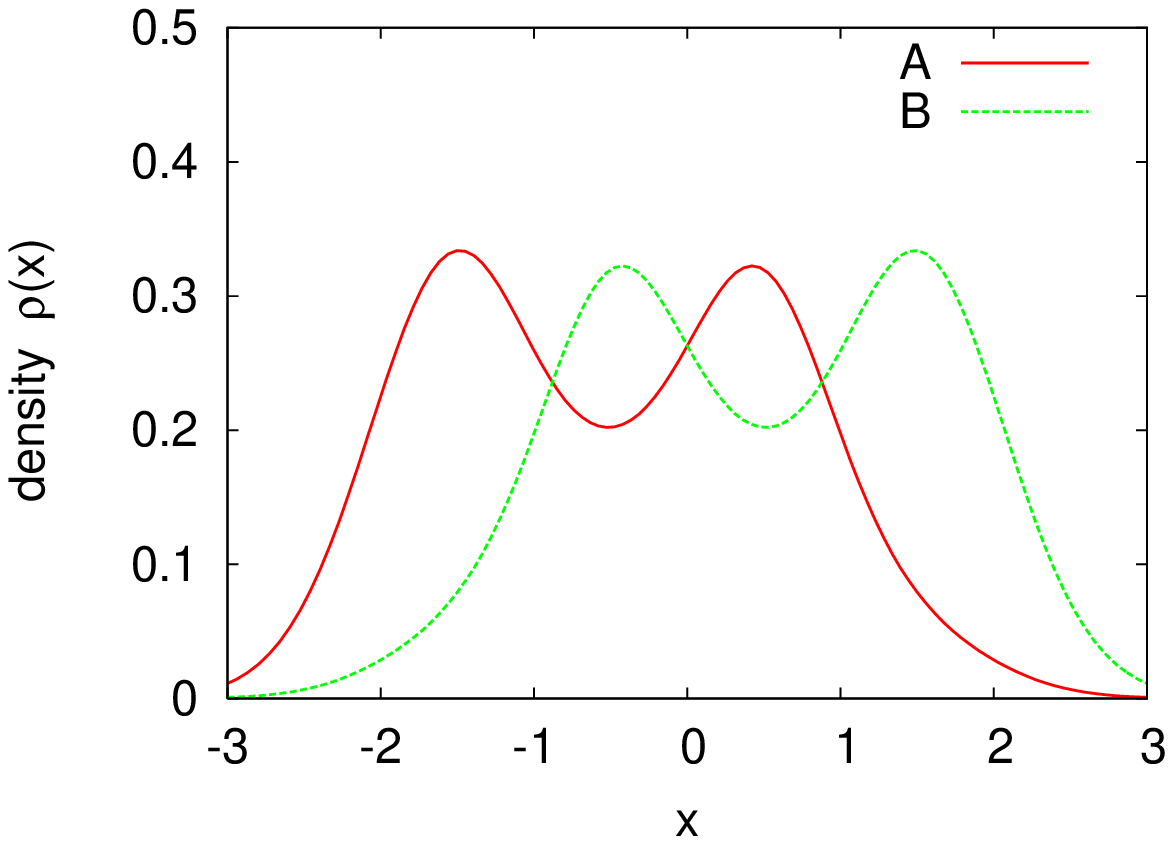}

\includegraphics[width=0.5\columnwidth,keepaspectratio]{figures/p2-2d1_DW\lyxdot 1\lyxdot 1\lyxdot 2\lyxdot h0_d}\includegraphics[width=0.5\columnwidth,keepaspectratio]{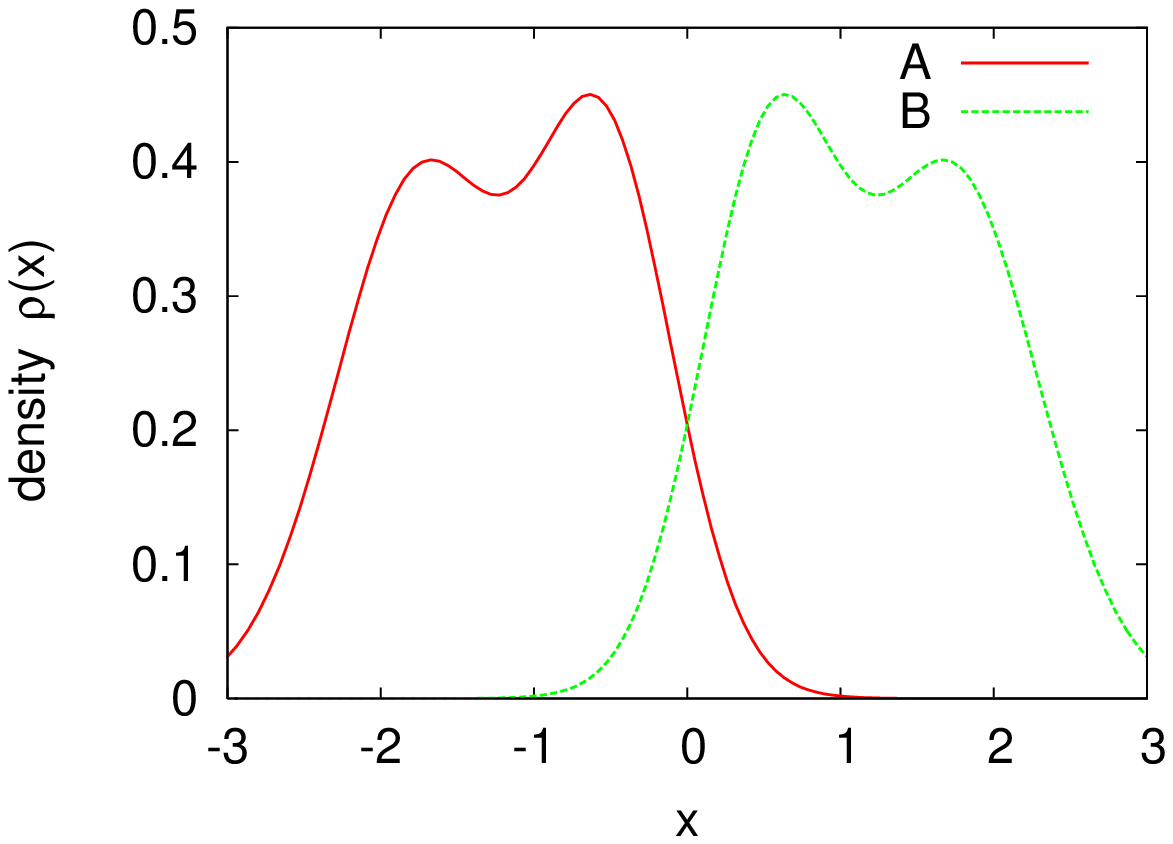}

\caption{(color online) Instability under symmetry-breaking perturbation,
$U_{\sigma}(x)=\frac{1}{2}x^{2}+d_{\sigma}x$ ($d_{\mathrm{A}}=-d_{\mathrm{B}}=0.1$).
\emph{Left}: Densities $\rho_{\sigma}(x)$ for $g_{\sigma}=0.4$;
\emph{Right}: $g_{\sigma}=25$. Shown are the coupling strengths $g_{\mathrm{AB}}=0.4,4.7,25$
from top to bottom.\label{cap:symbreak}}
\end{figure}

By permutation symmetry of $H$ between A and B, the density profiles
$\rho_{\sigma}$ are identical, even in the limit $g_{\mathrm{AB}}\gg g_{\sigma}$,
and thus trivially cannot exhibit phase separation. However, if we
deal with two different species, it is conceivable that these two
feel slightly different trapping potentials, where the deviations
are much weaker than the mean trapping and inter-particle forces but
only serve to break the symmetry. In particular, imagine that $U_{\sigma}(x)=\frac{1}{2}x^{2}+d_{\sigma}x$,
such that the trap centers be shifted by $d_{\mathrm{A}}=-d_{\mathrm{B}}\ll1$,
see Fig.~\ref{cap:symbreak}. Expectedly, for weak couplings $g_{\mathrm{AB}}$,
the profiles are barely affected. However, toward stronger inter-species
repulsion, this tiny perturbation is the last straw needed to make
the two phases demix completely. Similar results hold also for different
densities, $N_{\mathrm{A}}\neq N_{\mathrm{B}}$. This makes it even
more inviting to think of the symmetric profiles in Figs.~\ref{cap:2+2-g0.4}--\ref{cap:2+2-g25}
as averages over the equivalent configurations with A (B) being on
the left (right), and the other way around.

\section{Phase-separation scenarios \label{sec:demixing}}

So far, we have studied a completely symmetric setup, where only the
inter-species interactions were permitted to differ. This way, a symmetry-breaking
perturbation was needed to reveal the hidden phase separation. Although
not experimentally unrealistic, this scenario is somewhat artificial.
We now want to relax the above symmetry constraints step by step and
discuss the wealth of different demixing pathways if the two components
have different particle numbers (Sec.~\ref{sub:Density-demixing}),
different internal interaction strengths (Sec.~\ref{sub:Interaction-demixing}),
and have different masses and/or trap frequencies (Sec.~\ref{sub:Trap-demixing}).

\subsection{Density-assisted demixing \label{sub:Density-demixing}}

\begin{figure*}
\includegraphics[width=0.33\textwidth,keepaspectratio]{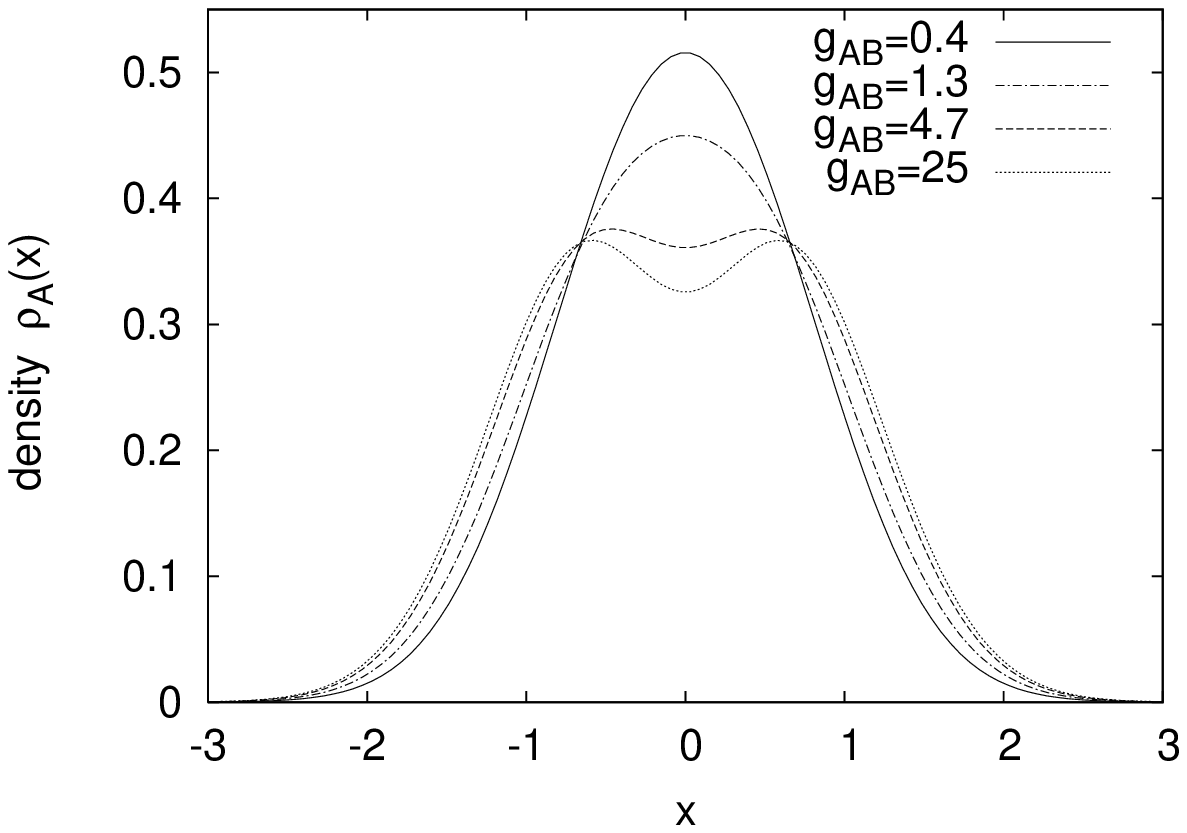}\includegraphics[width=0.33\textwidth,keepaspectratio]{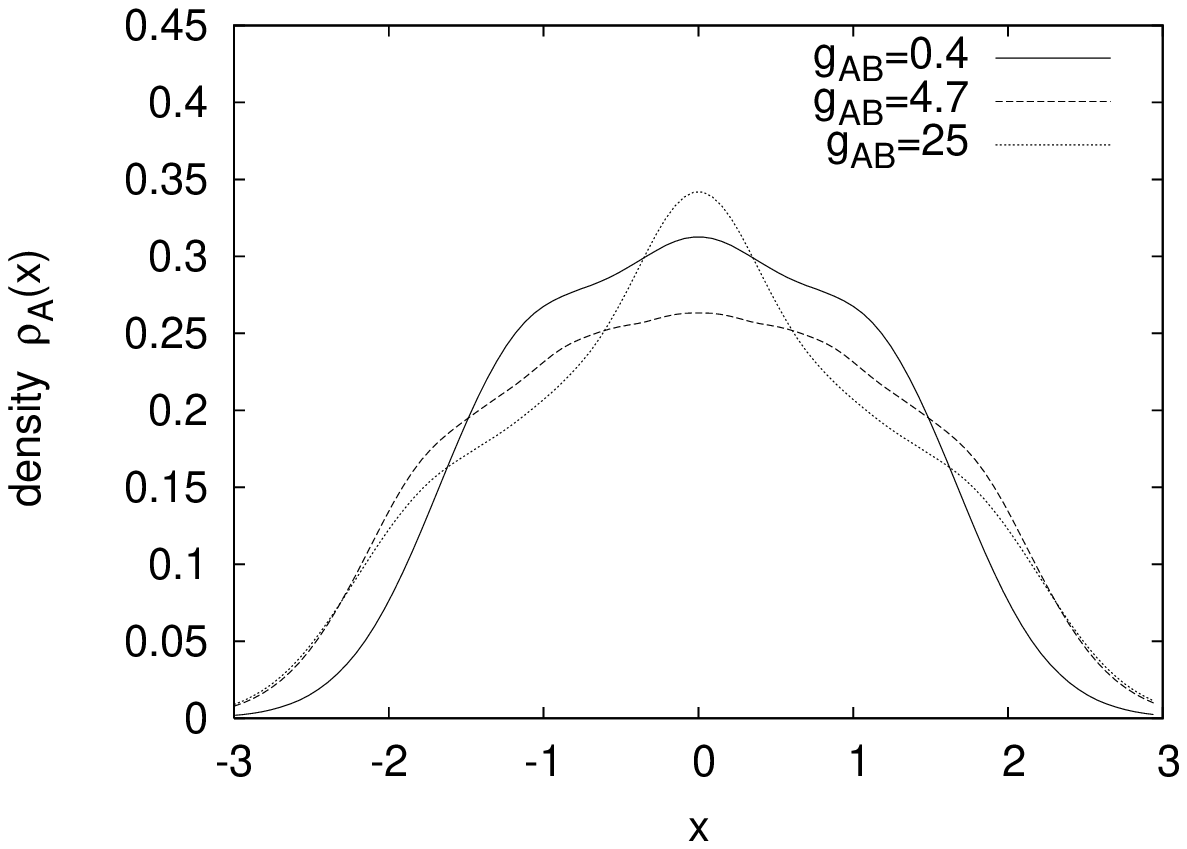}\includegraphics[width=0.33\textwidth,keepaspectratio]{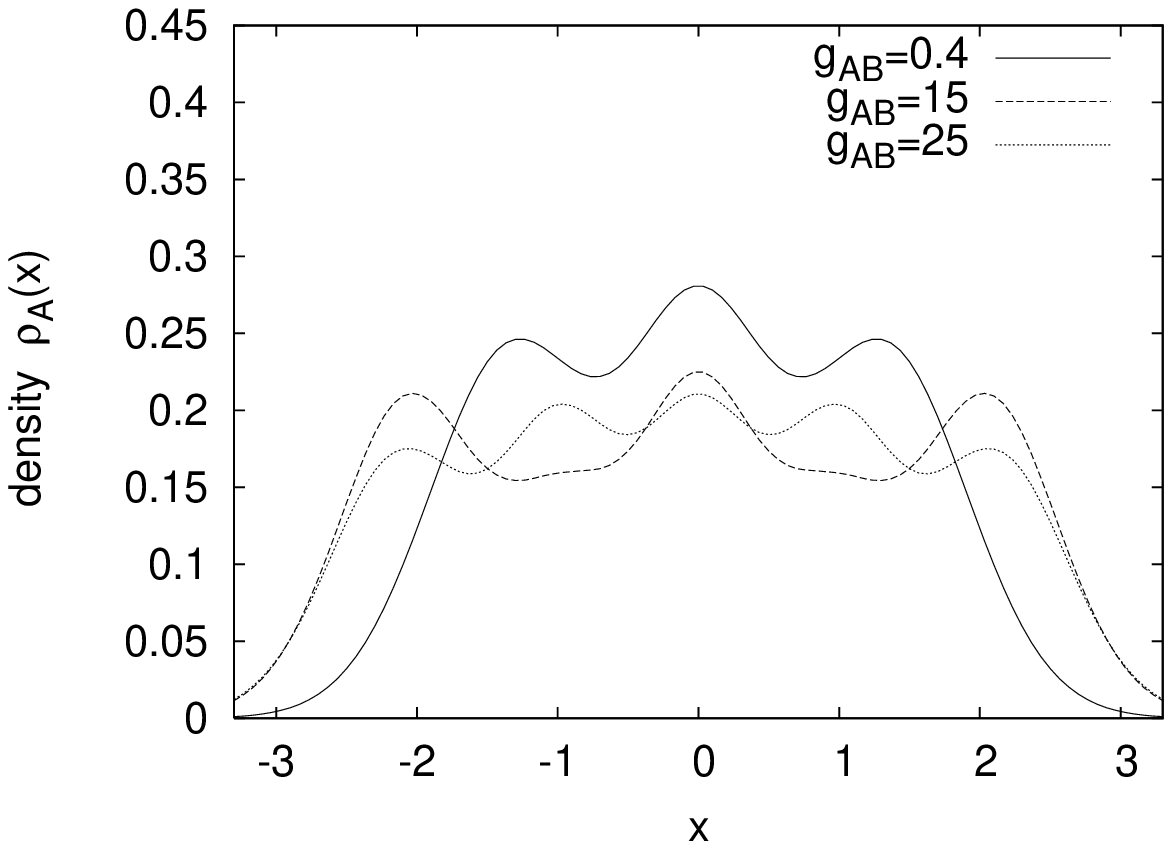}

\includegraphics[width=0.33\textwidth,keepaspectratio]{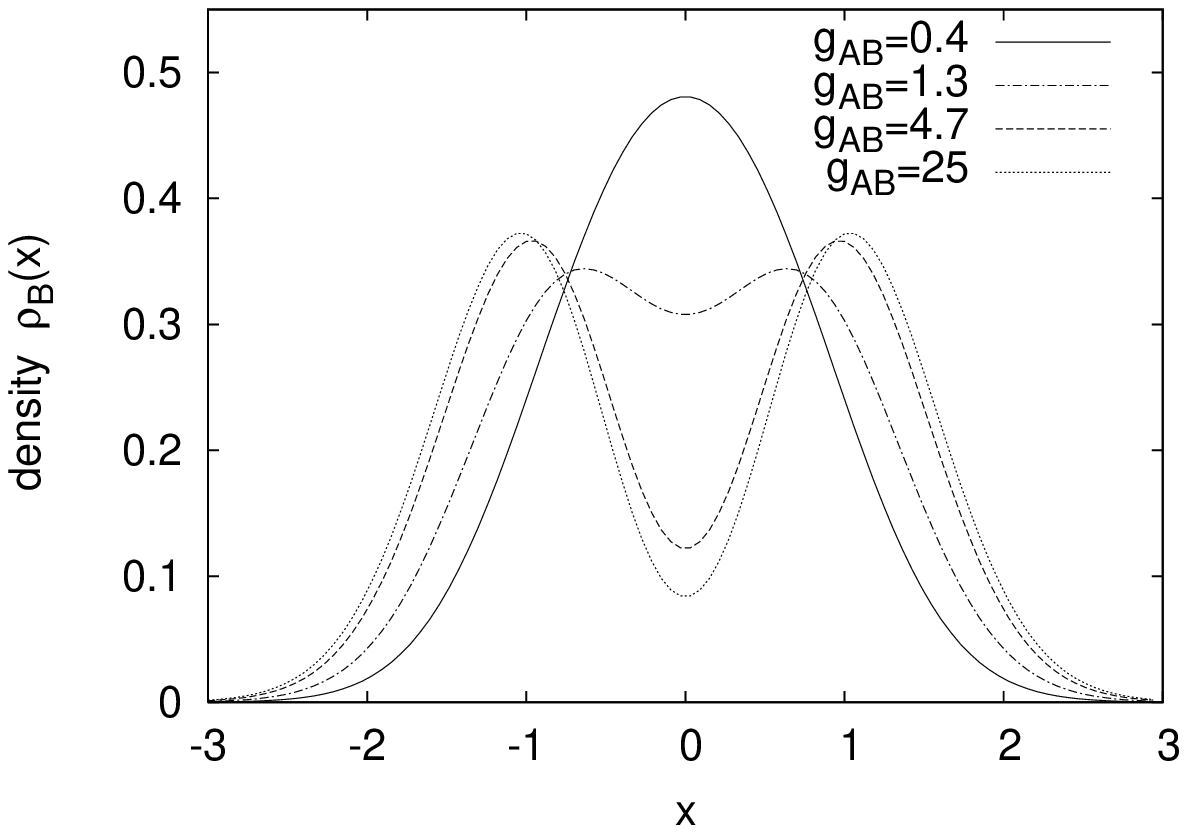}\includegraphics[width=0.33\textwidth,keepaspectratio]{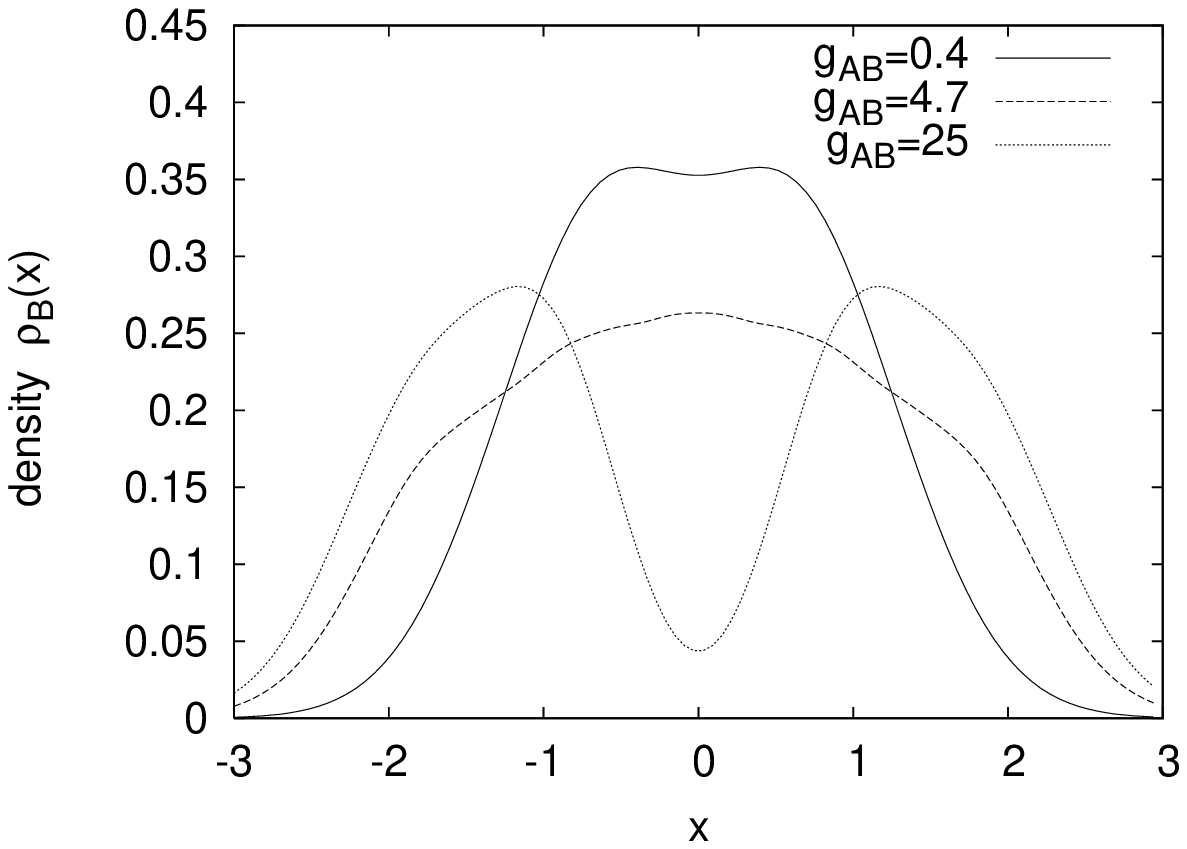}\includegraphics[width=0.33\textwidth,keepaspectratio]{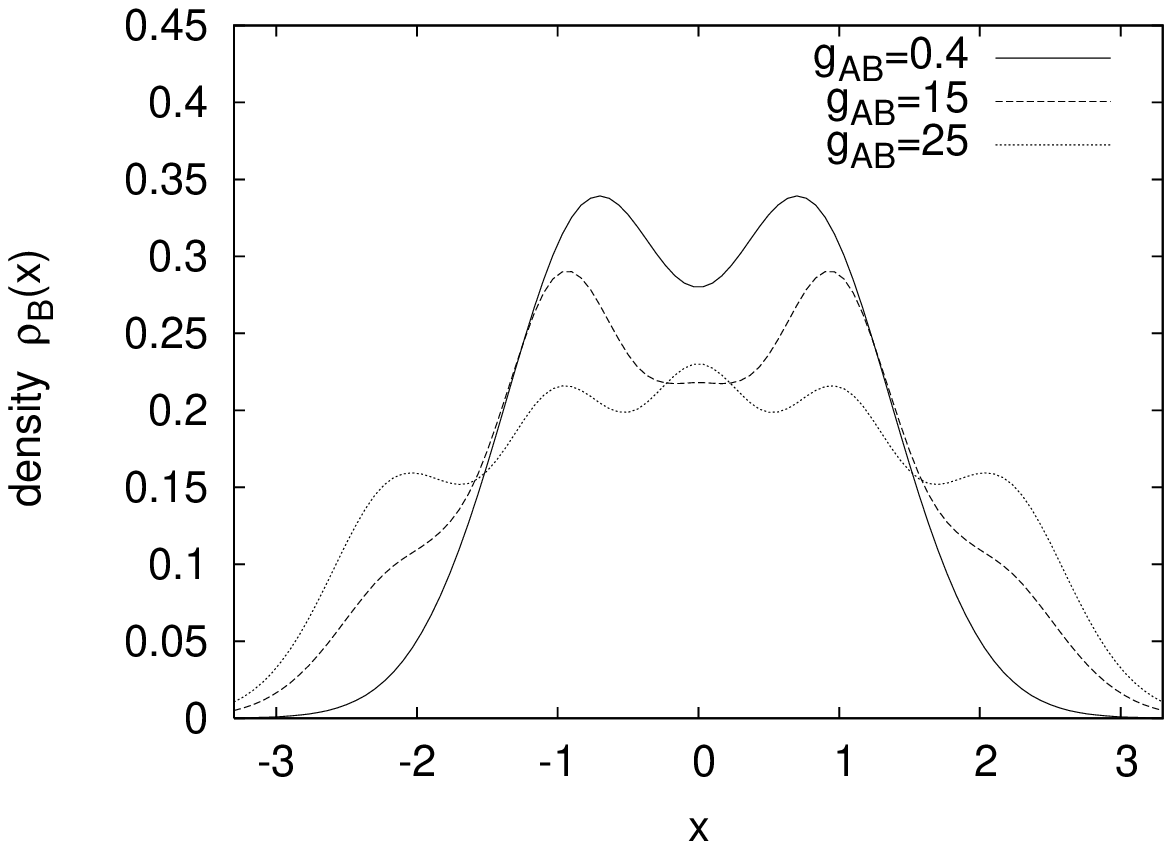}

\caption{Demixing for different particle numbers $N_{\mathrm{A}}=3$, $N_{\mathrm{B}}=2$:
Density profiles $\rho_{\mathrm{A}}(x)$ (top) and $\rho_{\mathrm{B}}(x)$
(bottom) for intra-species interaction strengths $g_{\sigma}=0.4,\,4.7$
and $25$ (from left to right). \label{cap:3+2-demix}}
\end{figure*}

An obvious question regarding our findings in the previous section
is: What happens in the case of unequal particle numbers, $N_{\mathrm{A}}\neq N_{\mathrm{B}}$?
Figure~\ref{cap:3+2-demix} illustrates this on the example of two
weakly interacting components, $g_{\sigma}=0.4$ (\emph{left column}),
where $N_{\mathrm{A}}=3$ is larger than $N_{\mathrm{B}}=2$. As the
Bose-Bose coupling gets stronger, $g_{\mathrm{AB}}=1.3$, one observes
that the low-density phase B moves to the outer edge, thus {}``sandwiching''
high-density A component in the middle. This well-known phenomenon
traces back to the fact that the coupling energy $\langle H_{\mathrm{AB}}\rangle=N_{\mathrm{A}}N_{\mathrm{B}}g_{\mathrm{AB}}\int\rho_{\mathrm{AB}}(x,x)dx$
scales with $N_{\mathrm{A}}N_{\mathrm{B}}$ compared to the individual
energies $\langle H_{\sigma}\rangle\propto N_{\sigma}$. Thus for
the smaller B component, it is less expensive to move to the higher
potential regions. Not unexpectedly, we have found this to be even
more pronounced for $N_{\mathrm{A}}\gg N_{\mathrm{B}}$. Note that,
close to the composite-fermionization limit ($g_{\mathrm{AB}}=4.7,\,25$),
both components develop two humps in the density, if much more pronounced
for B. This is indicative of a superposition state similar to that
in Sec.~\ref{sec:sym}: The B atoms are found on the right and the
A atoms on the left, and \emph{vice versa}, only that the shift for
A is much smaller due to their higher density.

A similar pathway again exists for non-negligible intra-component
interactions, $g_{\sigma}=4.7$ (Fig.~\ref{cap:3+2-demix}). As $g_{\mathrm{AB}}\to\infty$,
the initially mixed phases separate: The profile $\rho_{\mathrm{A}}(x)$
develops a clear-cut peak at $x=0$, whereas B is again driven to
the boundary. Even though, on the face of it, this looks different
from the weakly interacting case, this density pattern can be understood
in complete analogy: The two components are isolated on the left and
on the right, respectively; however, due to the larger atom number
$N_{\mathrm{A}}$ and the repulsion pressure, A tends to be more in
the center \emph{on average}.

An entirely different situation is encountered in the {}``Fermi-Fermi''-like
setup with $g_{\sigma}=25$ (Fig.~\ref{cap:3+2-demix}). For intermediate
$g_{\mathrm{AB}}=15<g_{\mathrm{\sigma}}$, a phase forms where the
A atoms localize at three discrete spots such that the two B atoms
fill the two holes in between. For $g_{\mathrm{AB}}=25$, by contrast,
the $N$ atoms completely localize atom by atom just like in the case
of a 2+2 mixture, in agreement with the extended Bose-Fermi map (\ref{eq:BFmap}).
In analogy to Sec.~\ref{sec:sym}, these will demix under slight
symmetry-breaking perturbations into one phase with $N_{\mathrm{A}}=3$
density wiggles on, say, the left side, and $N_{\mathrm{B}}=2$ on
the right.

\subsection{Interaction-assisted demixing \label{sub:Interaction-demixing}}

\begin{figure}
\begin{centering}\includegraphics[width=0.8\columnwidth,keepaspectratio]{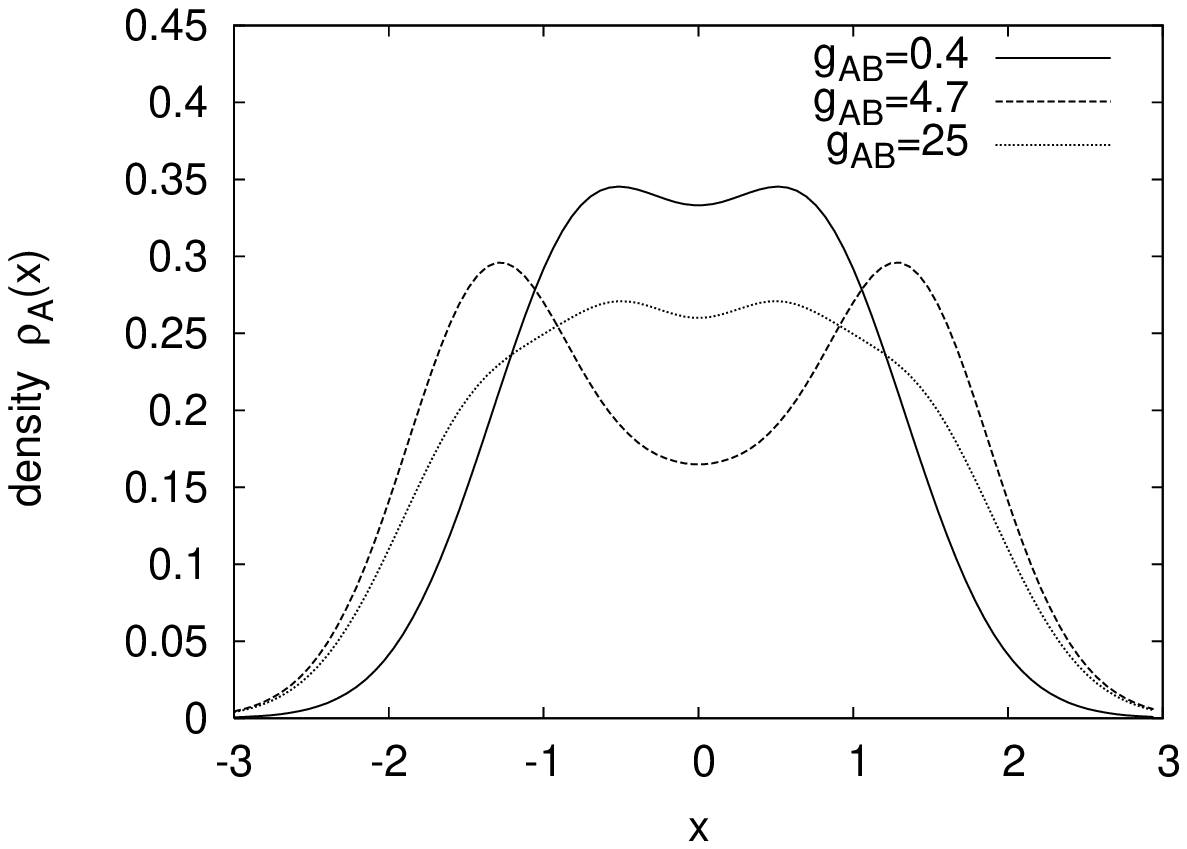}\par\end{centering}

\begin{centering}\includegraphics[width=0.8\columnwidth,keepaspectratio]{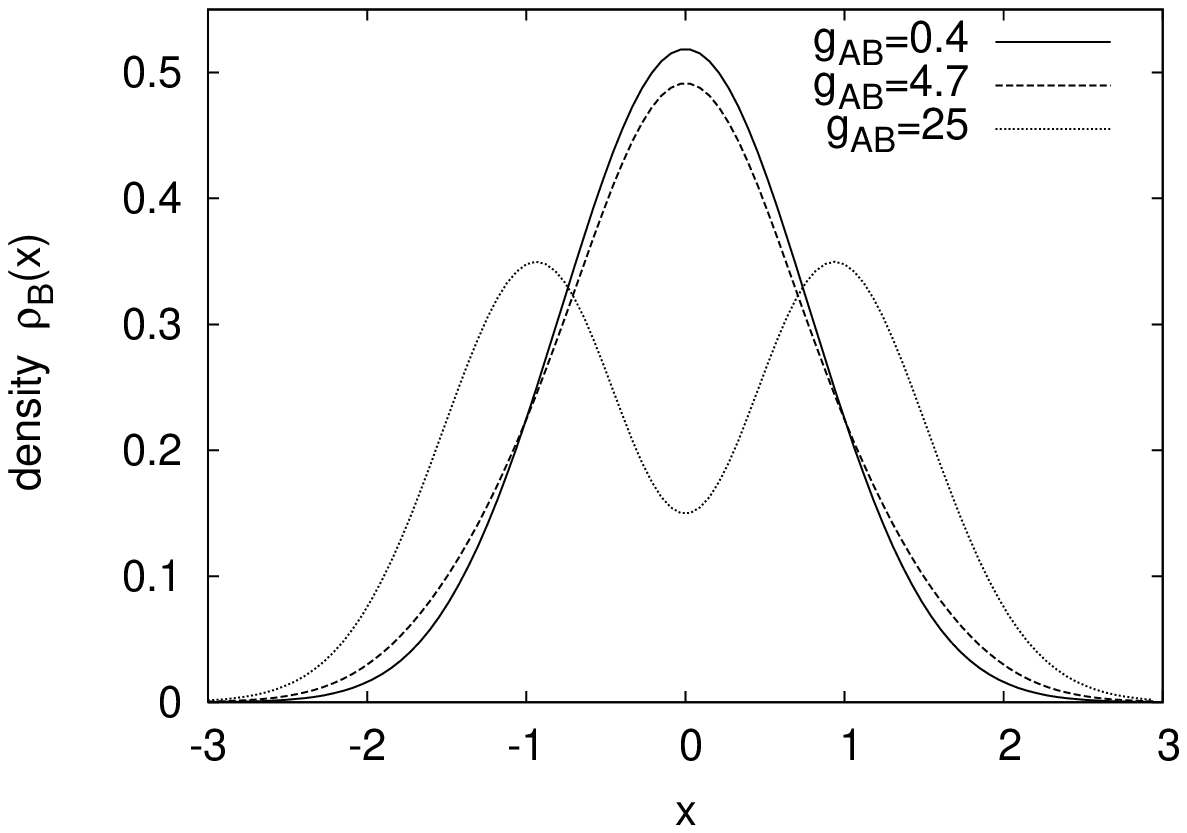}\par\end{centering}

\caption{Demixing for different intra-component interactions $g_{\mathrm{A}}=4.7$,
$g_{\mathrm{B}}=0.4$: Density profiles $\rho_{\mathrm{A}}(x)$ (top)
and $\rho_{\mathrm{B}}(x)$ (bottom) for an $N_{\sigma}=2$ mixture
at different couplings $g_{\mathrm{AB}}=0.4,\,4.7$ and $25$. \label{cap:2+2-demix}}
\end{figure}

Up until now, we have assumed comparable interactions within each
component. Of course, it is of fundamental concern what the composite-fermionization
crossover looks like in the case where one species is more strongly
repulsive, including as a special case a {}``Bose-Fermi''-type mixture
of one weakly interacting and another, fermionized component. 

An illustrative example is given in Fig.~\ref{cap:2+2-demix}, displaying
the composite-fermionization crossover for an $N_{\sigma}=2$ mixture
with $g_{\mathrm{A}}=4.7>g_{\mathrm{B}}=0.4$. We distinguish two
regimes: 

\begin{enumerate}
\item For $g_{\mathrm{AB}}<g_{\mathrm{A}}$, the weakly interacting central
B cloud is barely affected; at the same time, the strongly interacting
A bosons move slightly to the outside, thus cutting down on both intra-
and inter-species interaction energy. 
\item By contrast, for $g_{\mathrm{AB}}=25>g_{\mathrm{A}}$, this partial
separation is no longer enough: Now the B cloud splits up, signifying
the formation of the entangled state discussed before with $N_{\mathrm{A}}$
atoms on the left and $N_{\mathrm{B}}$ atoms on the right, and \emph{vice
versa}. Note that, owing to the strong inter-species repulsion in
A, the two humps in $\rho_{\mathrm{A}}$ are washed out strongly.
\end{enumerate}
We stress that only regime (1.) exists for a Bose-Fermi-type mixture,
i.e., where $g_{\mathrm{A}}\to\infty$: The minimum-energy state for
 infinitely large $g_{\mathrm{AB}}$ then has all B atoms in the center
and A on the edges.

\paragraph*{Coherence aspects.}

\begin{figure}
\begin{centering}\includegraphics[width=0.51\columnwidth,keepaspectratio]{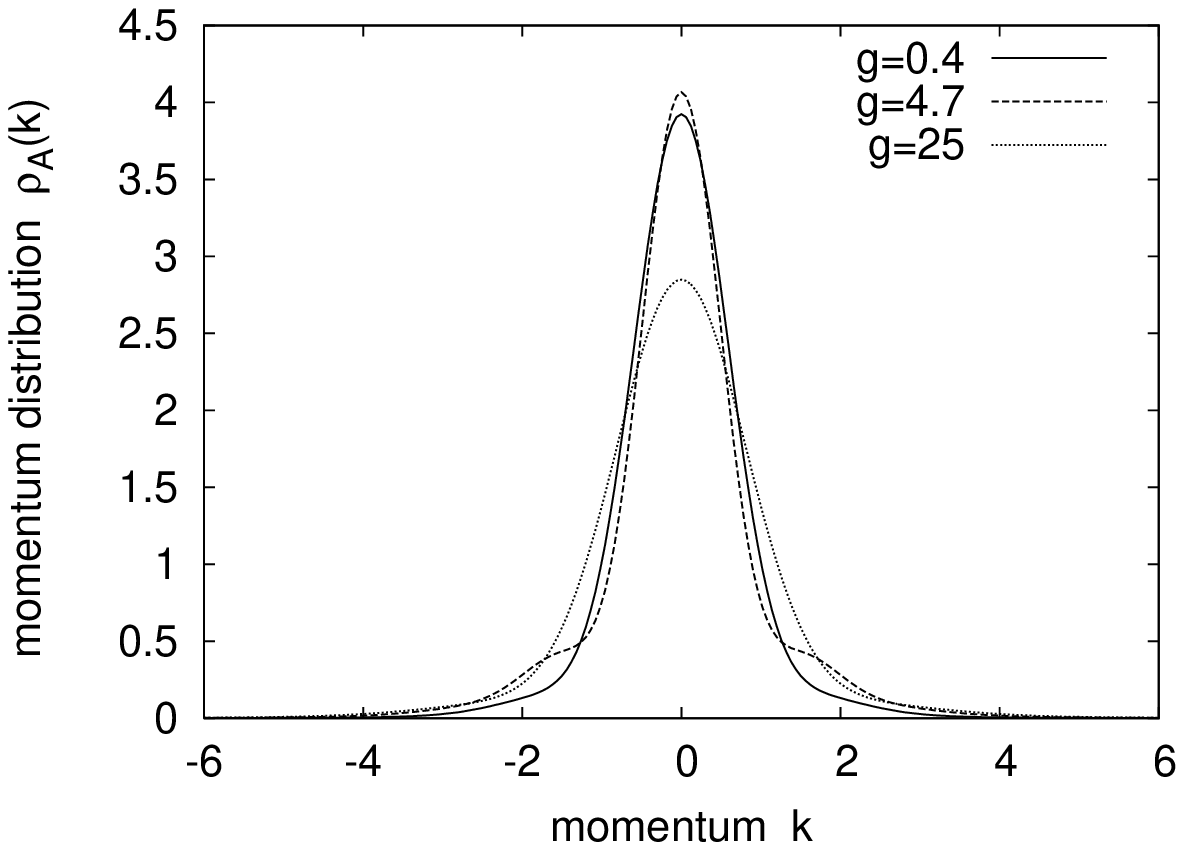}\includegraphics[width=0.51\columnwidth,keepaspectratio]{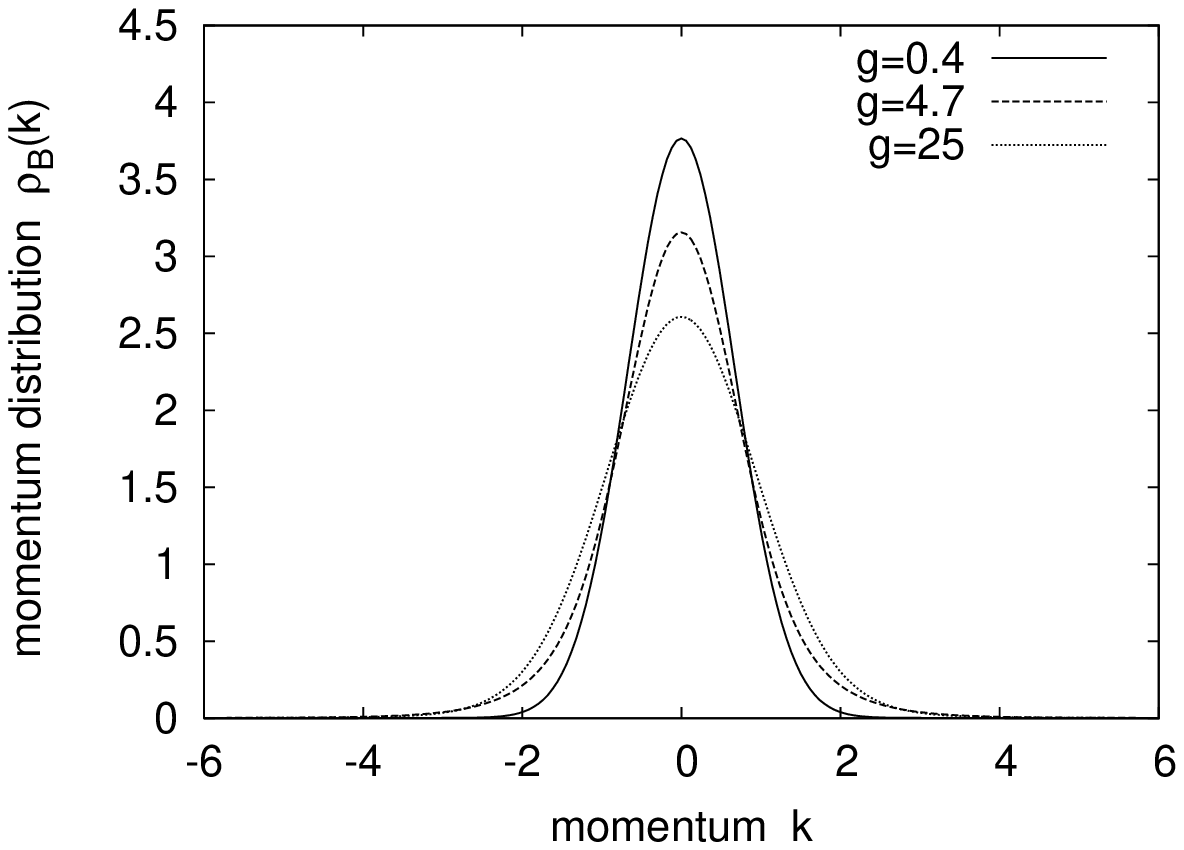}\par\end{centering}

\begin{centering}\includegraphics[width=0.3\columnwidth,keepaspectratio]{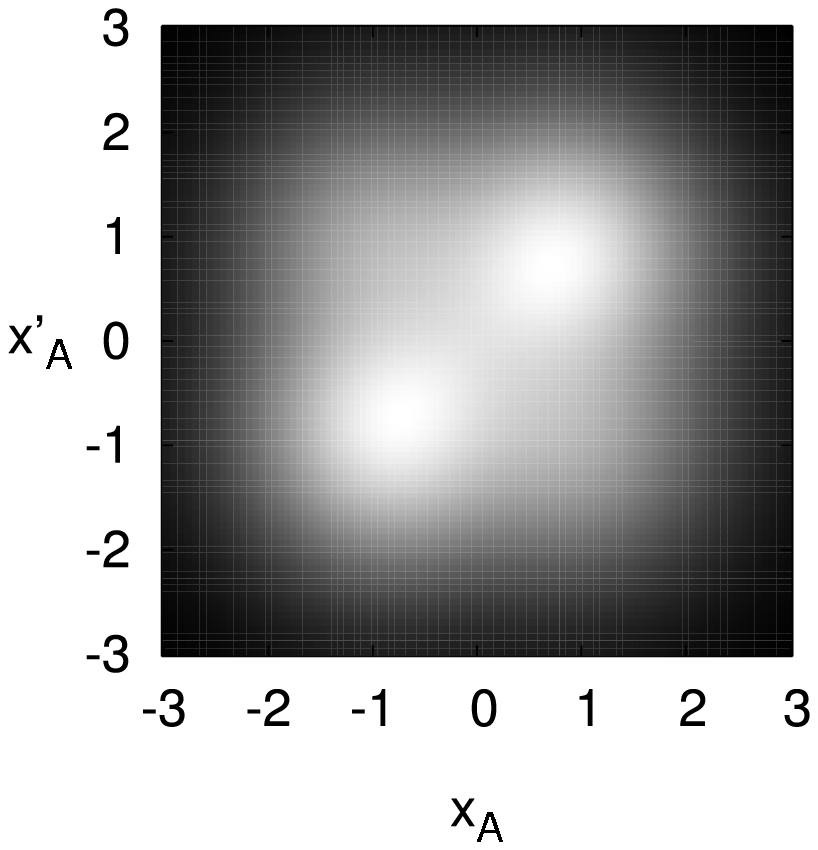}\includegraphics[width=0.3\columnwidth,keepaspectratio]{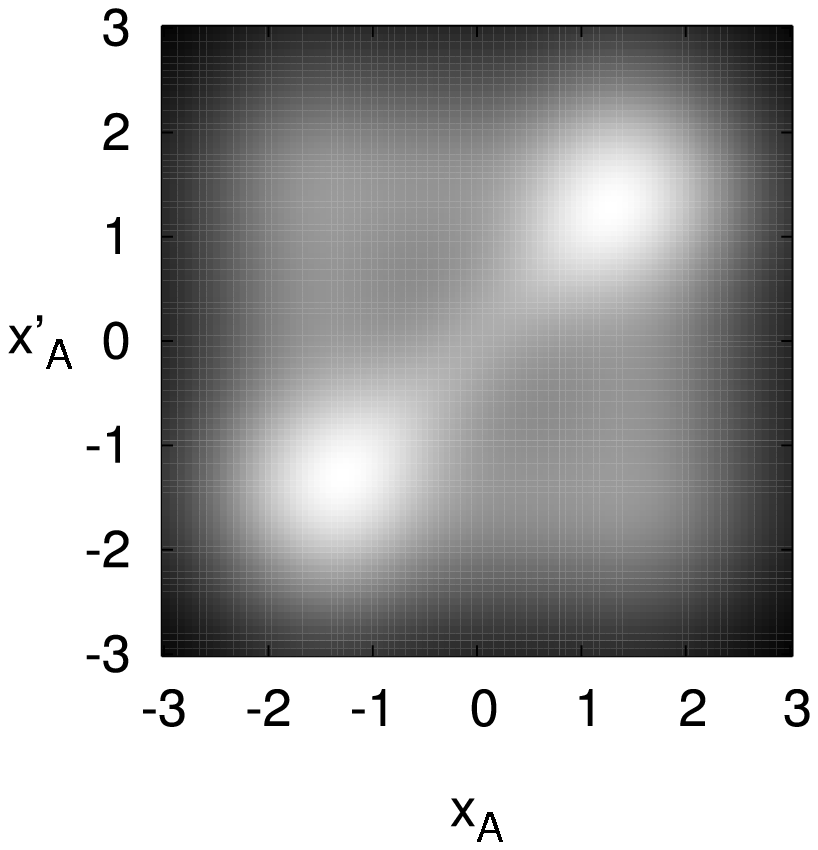}\includegraphics[width=0.3\columnwidth,keepaspectratio]{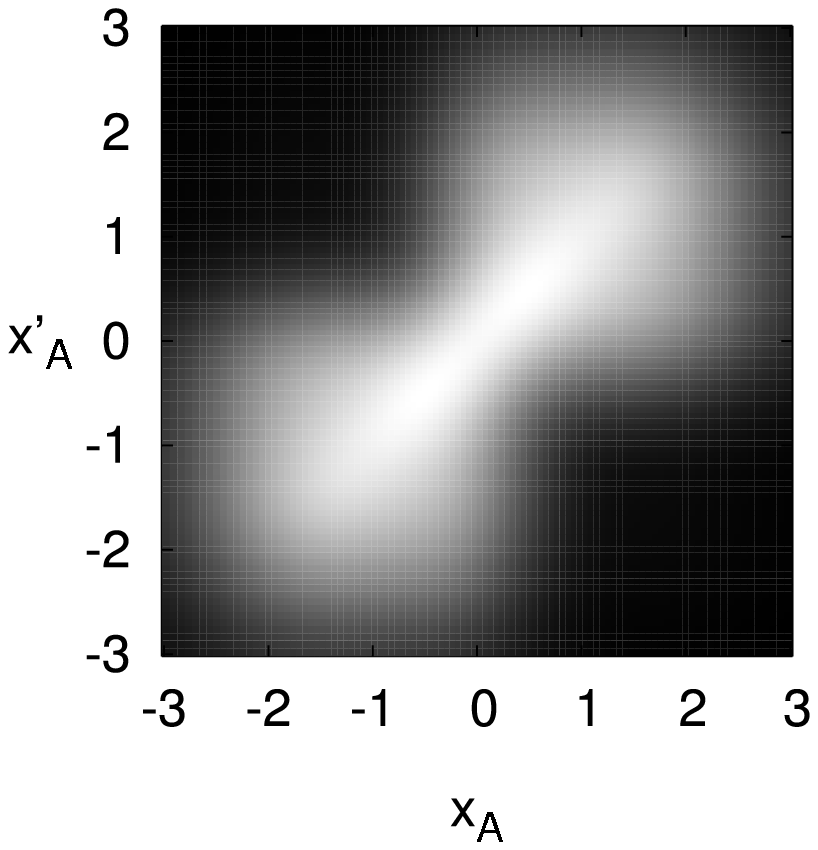}\par\end{centering}

\begin{centering}\includegraphics[width=0.3\columnwidth,keepaspectratio]{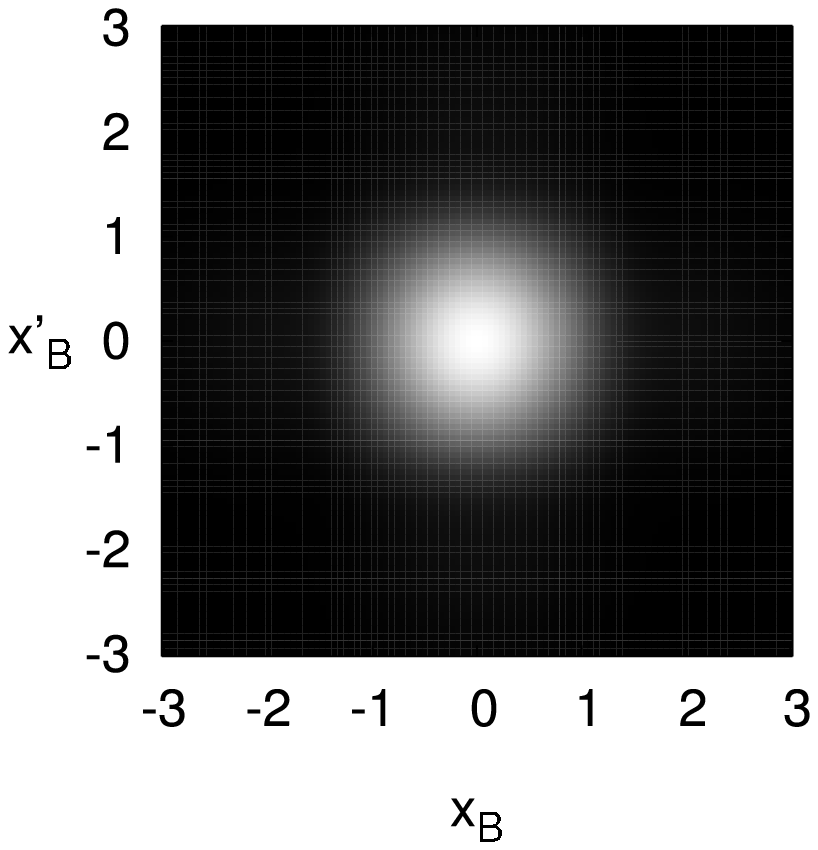}\includegraphics[width=0.3\columnwidth,keepaspectratio]{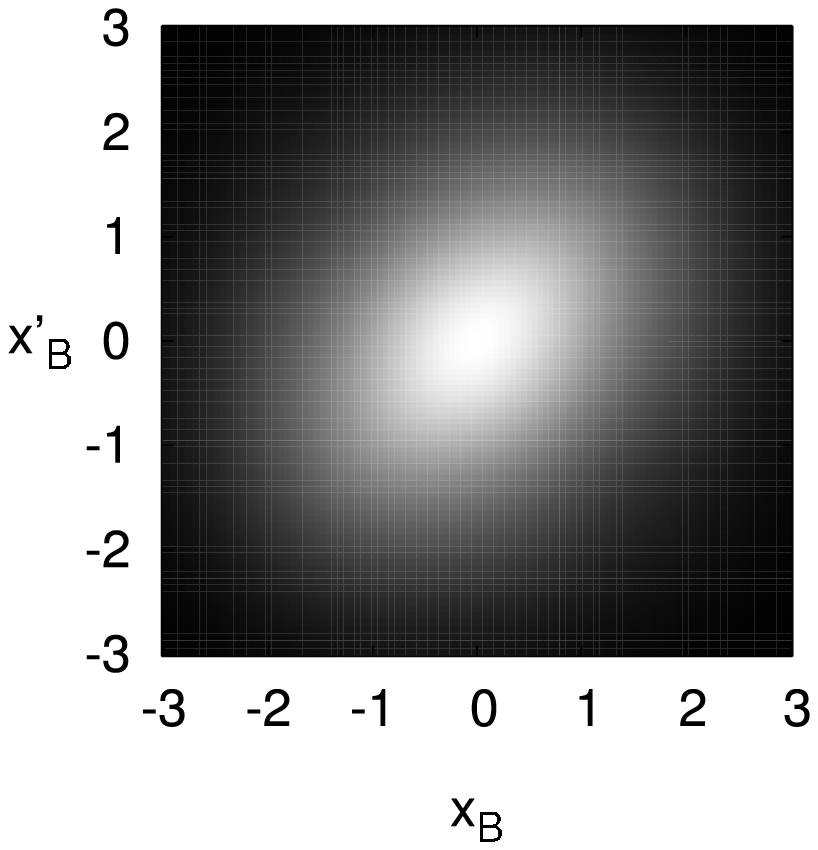}\includegraphics[width=0.3\columnwidth,keepaspectratio]{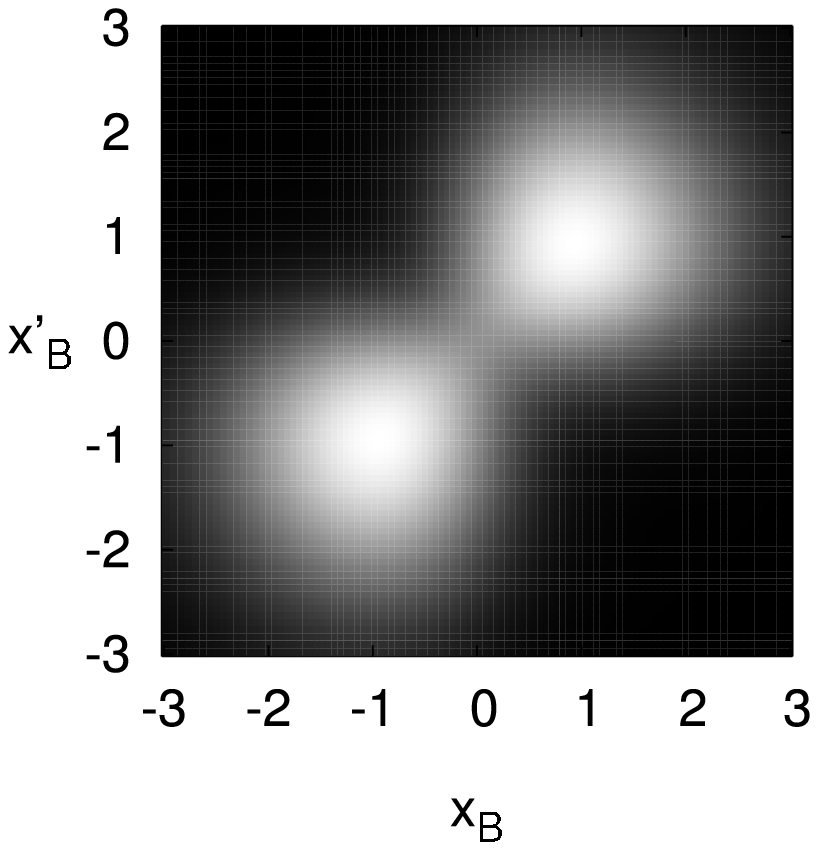}\par\end{centering}

\caption{Coherence properties of the system in Fig.~\ref{cap:2+2-demix}.
\emph{Top}: momentum distributions $\tilde{\rho}_{\sigma}(k)$; \emph{Bottom}:
One-body density matrices $\rho_{\mathrm{\sigma}}(x_{\sigma},x_{\sigma}')$
for $g_{\mathrm{AB}}=0.4,\,4.7$ and $25$ (from left to right). \label{cap:coherence}}
\end{figure}

At this point, it is worthwhile dwelling for a moment on the coherence
properties of bosonic mixtures, as reflected in the reduced one-body
density matrix $\rho_{\mathrm{\sigma}}(x,x')\equiv\langle x|\hat{\rho}_{\sigma}^{(1)}|x'\rangle$
and, closely related, the momentum distribution \[
\tilde{\rho}_{\sigma}(k)\equiv2\pi\langle k|\hat{\rho}_{\sigma}^{(1)}|k\rangle=\int\negmedspace dx\int\negmedspace dx'e^{-ik(x-x')}\rho_{\sigma}(x,x').\]
It has been demonstrated for \emph{identical} bosons \cite{zoellner06b,deuretzbacher06}
how, in the course of fermionization, the zero-momentum peak $\tilde{\rho}(k=0)$---related
to the fraction of condensed bosons---is attenuated and redistributed
toward higher momenta, culminating in a characteristic decay $\tilde{\rho}(k)\stackrel{k\to\infty}{\sim}ck^{-4}$
as predicted for hard-core short-range interactions \cite{minguzzi02}.
Equivalently, the off-diagonal long-range order, measured by $\rho_{1}(x,-x)$
as $x\to\infty$, is strongly reduced. We generally find the same
two mechanisms at work here, which we exemplify in Fig.~\ref{cap:coherence}.
For stronger inter-species repulsion, the high-momentum tail in $\tilde{\rho}_{\sigma}(k)$
becomes more pronounced. Interestingly, for the component B with weaker
interaction, the $k=0$ peak starts diminishing right away, while
the strongly repulsive A component first sees a sharpening at zero
momentum ($g_{\mathrm{AB}}=4.7$). This derives from its initial delocalization
so as to move away from B {[}cf. $\rho_{\mathrm{A}}(x_{\mathrm{A}},x'_{\mathrm{A}})$
in Fig.~\ref{cap:coherence}], which allows the A atoms to spend
less kinetic energy $\langle\frac{1}{2}p^{2}\rangle\propto(\Delta{k)}^{2}$. 

\subsection{Trap-induced demixing \label{sub:Trap-demixing}}

After having explored the effect of different densities or interaction
strengths on the composite-fermionization pathway, let us now relax
the condition of equal masses and trapping potentials (\emph{here}:
frequencies). In this case, the system no longer maps to a single-component
Bose gas even for $g_{\sigma}=g_{\mathrm{AB}}$.

\subsubsection{Different confinement lengths}

\begin{figure}
\begin{centering}\includegraphics[width=0.8\columnwidth,keepaspectratio]{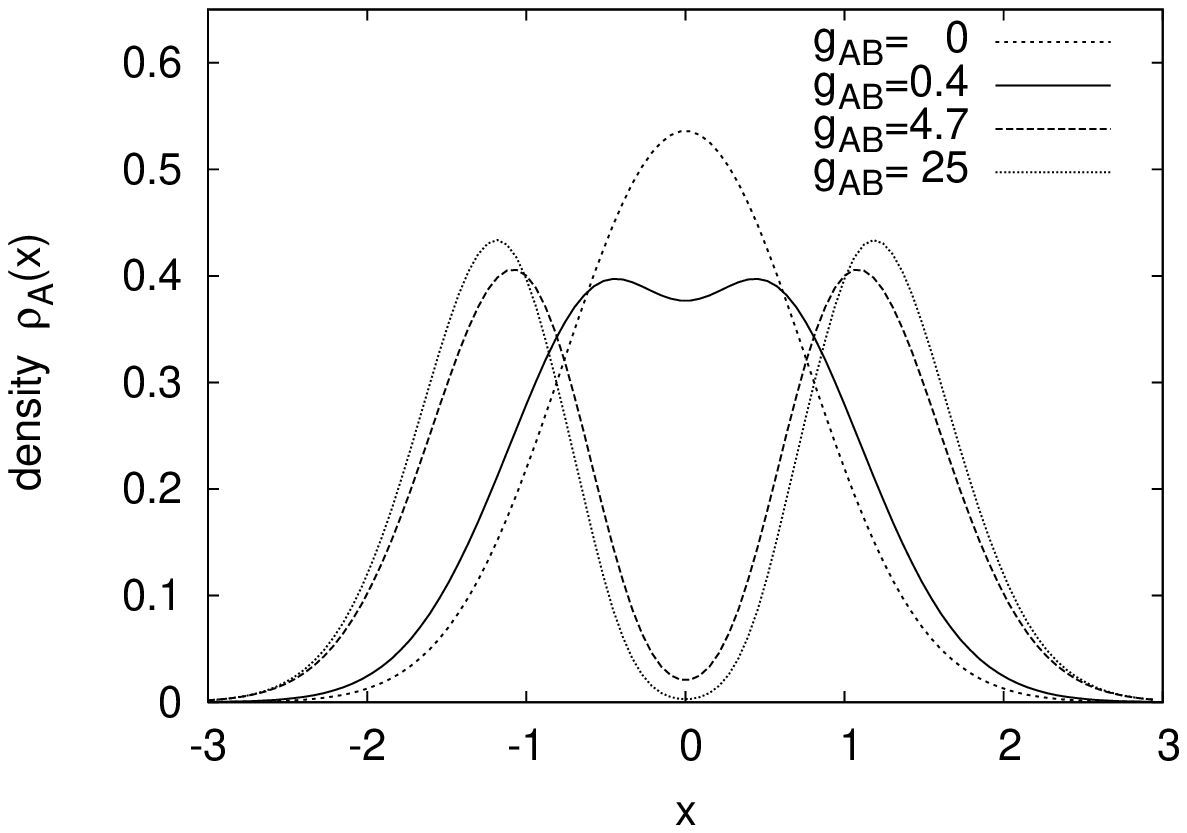}\par\end{centering}

\begin{centering}\includegraphics[width=0.8\columnwidth,keepaspectratio]{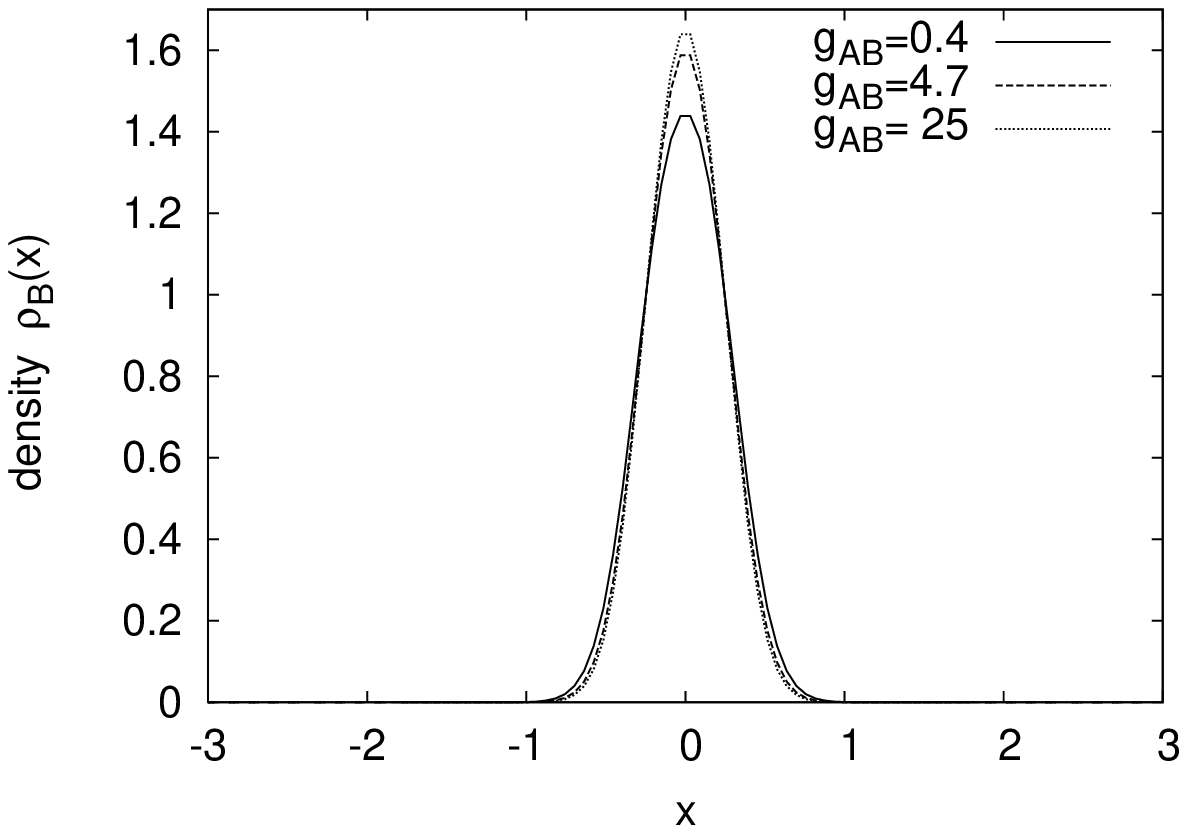}\par\end{centering}

\caption{Demixing for different masses, $M_{\mathrm{B}}/M_{\mathrm{A}}=9$:
Density profiles $\rho_{\mathrm{A}}(x)$ (top) and $\rho_{\mathrm{B}}(x)$
(bottom) for a $2+2$ mixture with $g_{\mathrm{\sigma}}=0.4$ at different
couplings $g_{\mathrm{AB}}$. \label{cap:2+2_m9}}
\end{figure}

Assume that we have a nontrivial mass ratio, i.e., $M_{\mathrm{B}}>1$
without loss of generality, with an otherwise symmetric parameter
set. The effective oscillator length of the B atoms will then be reduced
by a factor of $a_{\mathrm{B}}=1/\sqrt{M_{\mathrm{B}}}<1$. This situation
is visualized in Fig.~\ref{cap:2+2_m9} for the choice $M_{\mathrm{B}}=9$.
At weak couplings, $\rho_{\mathrm{B}}(x)$ is simply constricted at
the trap center, while $\rho_{\mathrm{A}}(x)$ extends over a much
larger region. As we switch on the interaction between the components,
the B atoms remain unmoved, whereas the A bosons are gradually driven
toward the outside. This is intuitive: The former component roughly
feels an average Hamiltonian \[
\bar{H}_{\mathrm{B}}=H_{\mathrm{B}}+\mathrm{tr_{A}}[H_{\mathrm{AB}}\hat{\rho}_{\mathrm{A}}^{(N_{\mathrm{A}})}]=H_{\mathrm{B}}+g_{\mathrm{AB}}N_{\mathrm{A}}\sum_{b=1}^{N_{\mathrm{B}}}\rho_{\mathrm{A}}(x_{\mathrm{B},b}),\]
and likewise for A. Since the heavy B atoms are effectively frozen
at the center, where $\rho_{\mathrm{A}}(x_{\mathrm{B}})\approx\rho_{\mathrm{A}}(0)$
changes slowly, they only feel a constant energy shift due to the
presence of A atoms. By contrast, the latter ones see an effective
{}``potential barrier'' $\rho_{\mathrm{B}}(x_{\mathrm{A}})\approx\delta(x_{\mathrm{A}})$
which varies only in a small region about zero. 

That phase-separation mechanism is largely insensitive to the intra-species
interactions $g_{\sigma}$: We have confirmed these results also for,
e.g., two quasi-fermionized components. Also note that a similar scale
separation persist for the case of different frequencies but equal
masses, i.e., $\omega_{\mathrm{B}}/\omega_{\mathrm{A}}\gg1$. The
different effective interaction felt by B, $g_{\mathrm{B}}'=g_{\mathrm{B}}\sqrt{M_{\mathrm{B}}/\omega_{\mathrm{B}}}$,
and the modified energy scale, $\omega_{\mathrm{B}}$, do not qualitatively
alter the picture above.

\subsubsection{Different energy scales}

\begin{figure}
\begin{centering}\includegraphics[width=0.8\columnwidth,keepaspectratio]{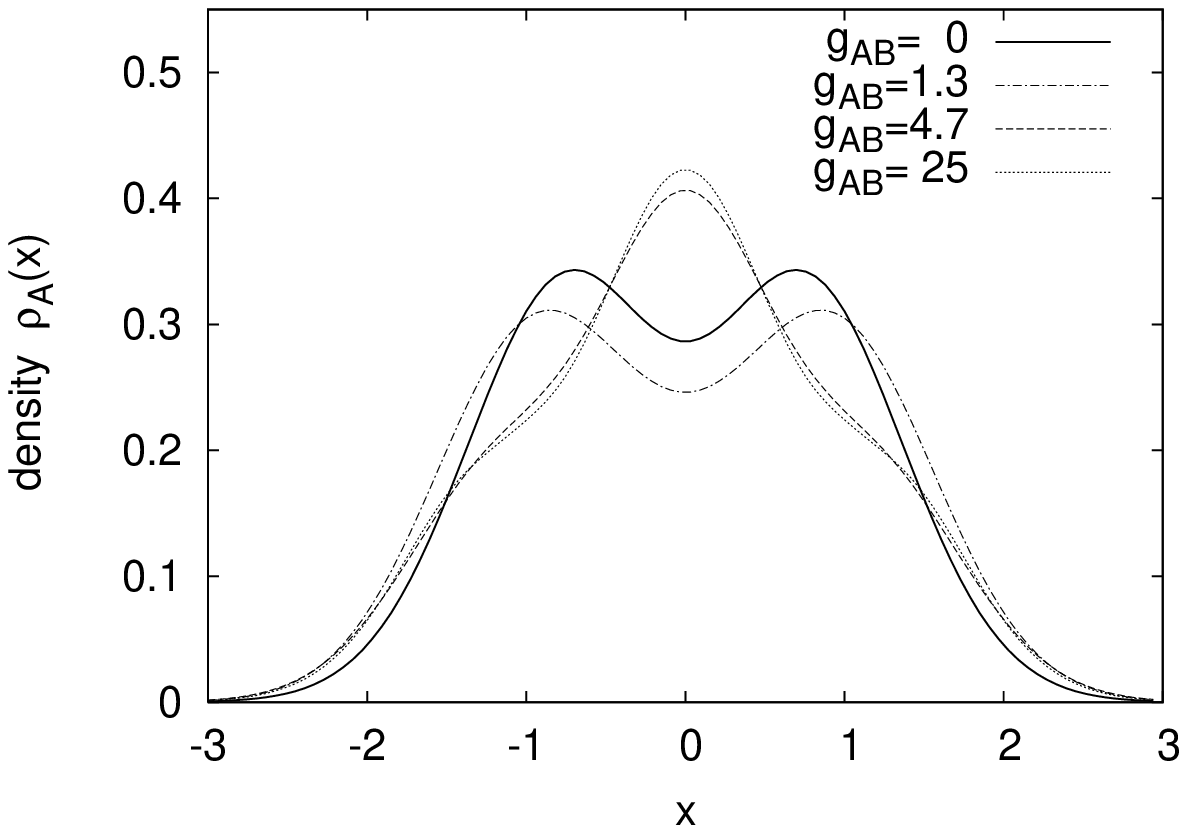}\par\end{centering}

\begin{centering}\includegraphics[width=0.8\columnwidth,keepaspectratio]{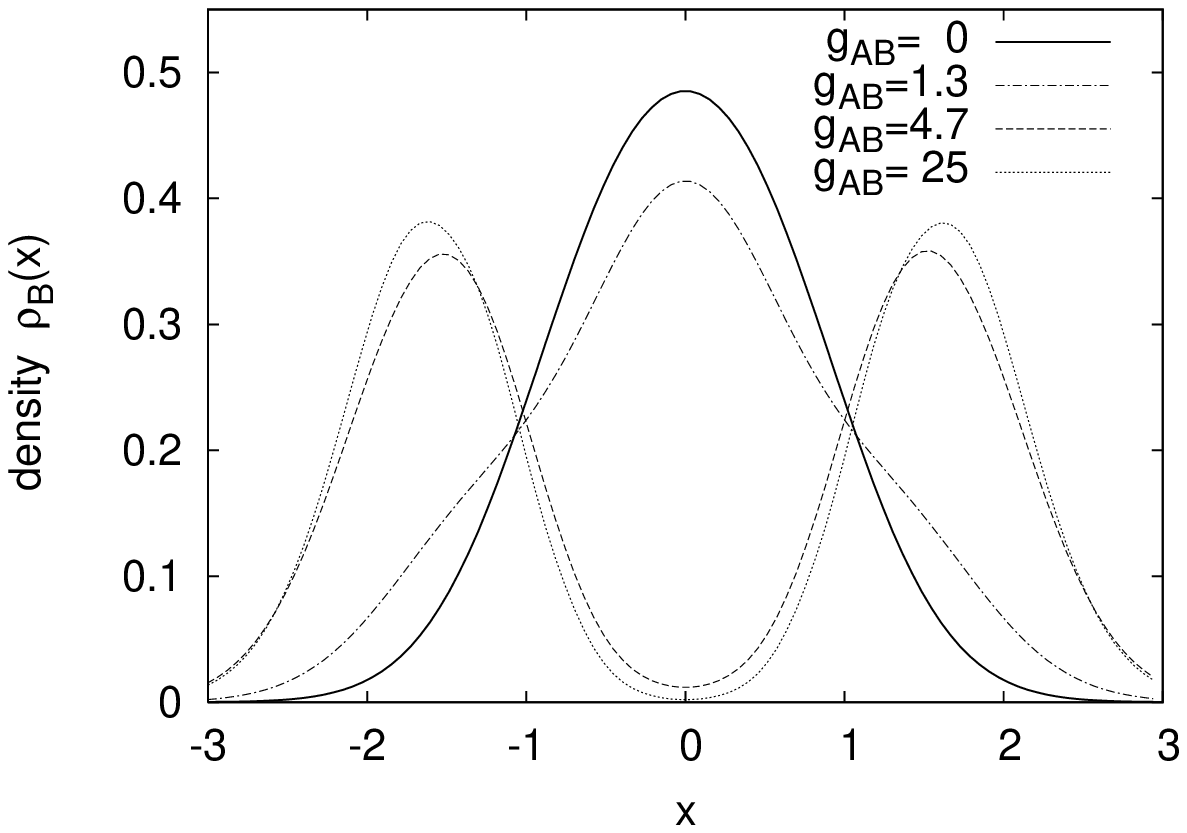}\par\end{centering}

\caption{Demixing for different energy scales, $M_{\mathrm{B}}=1/\omega_{\mathrm{B}}=3$:
Density profiles $\rho_{\mathrm{A}}(x)$ (top) and $\rho_{\mathrm{B}}(x)$
(bottom) for a $2+2$ mixture with $g_{\mathrm{A}}=25$ and $g_{\mathrm{B}}=0.4$,
at different couplings $g_{\mathrm{AB}}$. \label{cap:2+2_m3w.1}}
\end{figure}

Let us now look into the complementary case where the oscillator lengths
be equal, $a_{\mathrm{B}}=1/\sqrt{M_{\mathrm{B}}\omega_{\mathrm{B}}}=1$,
but such that the energy scale $\omega_{\mathrm{B}}\neq1$ shall be
different. In order words, a stronger localization by virtue of a
larger mass is compensated by a shallower trap for the B species.
This option is sounded out in Fig.~\ref{cap:2+2_m3w.1}, where $M_{\mathrm{B}}=3=\omega_{\mathrm{B}}^{-1}$.
The two profiles still overlap for weak couplings $g_{\mathrm{AB}}=0,\,1.3$.
For sufficiently strong inter-species repulsion, though, it apparently
becomes beneficial for the B atoms to spread out to larger $x$ so
as to segregate from A, which in turn is compressed on the inside.
This is particularly striking in the setup captured in Fig.~\ref{cap:2+2_m3w.1}:
Here the A component is squeezed even though it is fermionized and
thus possesses a high internal pressure. The reason for that counter-intuitive
behavior is simply that the potential-energy costs for the B phase
are lower by $\omega_{\mathrm{B}}=1/3$. 

\section{Conclusion}

We have studied binary few-boson mixtures in a one-dimensional harmonic
trap throughout the crossover from weak coupling to strong inter-component
repulsion. Depending on the intra-species interactions, different
pathways to a new kind of {}``composite fermionization'' have shown
up: For two weakly interacting Bose gases, the two phases segregate
as a whole, where the demixing for equal densities is obscured by
symmetry-induced entanglement fragile to displacement of the trap.
By contrast, for two strongly repulsive components, demixing occurs
on the atomic level. 

If one component has a lower density, then it tends to delocalize
toward the outer edge, while the high-density phase is compressed
in the center. Furthermore, in case one component is far more repulsive,
the crossover exhibits an intermediate regime where that species forms
a shell around the central, weakly interacting, component; only for
large inter-species couplings do they fully segregate. This is accompanied
by an increase (decrease) of the central momentum peak for the strongly
(weakly) interacting species. Finally, for different mass or frequency
ratios, one component freezes at the trap center, such that it acts
as an effective potential barrier for the more mobile species.

The small mixtures of strongly repulsive atoms studied here should
be experimentally accessible. The preparation and detection techniques
required are similar to those already available for few bosons of
a single species. The interaction forces may be tuned independently
over a wide range by varying the (inter- and intra-species) scattering
lengths as well as the transverse confinement, which parametrically
modifies the effective one-dimensional coupling strengths.

\begin{acknowledgments}
Financial support from the Landesstiftung Baden-Württemberg in the
framework of the project {}`Mesoscopics and atom optics of small
ensembles of ultracold atoms' is acknowledged by P.S. and S.Z.

\end{acknowledgments}
\bibliographystyle{prsty}
\bibliography{/home/sascha/paper/pra/DW/phd,/home/sascha/bib/mctdh}

\begin{thebibliography}{10}

\bibitem{pethick}
C.~J. Pethick and H. Smith, {\em Bose-Einstein condensation in dilute gases}
  (Cambridge University Press, Cambridge, 2001).

\bibitem{pitaevskii}
L. Pitaevskii and S. Stringari, {\em Bose-Einstein Condensation} (Oxford
  University Press, Oxford, 2003).

\bibitem{truscott01}
A.~G. Truscott {\it et~al.}, Science {\bf 291},  2570  (2001).

\bibitem{hadzibabic02}
Z. Hadzibabic {\it et~al.}, Phys. Rev. Lett. {\bf 88},  160401  (2002).

\bibitem{taglieber08}
M. Taglieber {\it et~al.}, Phys. Rev. Lett. {\bf 100},  010401  (2008).

\bibitem{myatt97}
C.~J. Myatt {\it et~al.}, Phys. Rev. Lett. {\bf 78},  586  (1997).

\bibitem{hall98}
D.~S. Hall {\it et~al.}, Phys. Rev. Lett. {\bf 81},  1539  (1998).

\bibitem{maddaloni00}
P. Maddaloni {\it et~al.}, Phys. Rev. Lett. {\bf 85},  2413  (2000).

\bibitem{modugno02}
G. Modugno {\it et~al.}, Phys. Rev. Lett. {\bf 89},  190404  (2002).

\bibitem{catani08}
J. Catani {\it et~al.}, Phys. Rev. A {\bf 77},  011603  (2008).

\bibitem{cazalilla03}
M.~A. Cazalilla and A.~F. Ho, Phys. Rev. Lett. {\bf 91},  150403  (2003).

\bibitem{li03}
Y.-Q. Li, S.-J. Gu, Z.-J. Ying, and U. Eckern, Europhys. Lett. {\bf 61},  368
  (2003).

\bibitem{shchesnovich04}
V.~S. Shchesnovich, A.~M. Kamchatnov, and R.~A. Kraenkel, Phys. Rev. A {\bf
  69},  033601  (2004).

\bibitem{alon06}
O.~E. Alon, A.~I. Streltsov, and L.~S. Cederbaum, Phys. Rev. Lett. {\bf 97},
  230403  (2006).

\bibitem{mishra07}
T. Mishra, R.~V. Pai, and B.~P. Das, Phys. Rev. A {\bf 76},  013604  (2007).

\bibitem{roscilde07}
T. Roscilde and J.~I. Cirac, Phys. Rev. Lett. {\bf 98},  190402  (2007).

\bibitem{nho07}
K. Nho and D.~P. Landau, Phys. Rev. A {\bf 76},  053610  (2007).

\bibitem{kleine08}
A. Kleine {\it et~al.}, Phys. Rev. A {\bf 77},  013607  (2008).

\bibitem{koehler06}
T. K\"{o}hler, K. G\'{o}ral, and P.~S. Julienne, Rev. Mod. Phys. {\bf 78},
  1311  (2006).

\bibitem{Olshanii1998a}
M. Olshanii, Phys. Rev. Lett. {\bf 81},  938  (1998).

\bibitem{kinoshita04}
T. Kinoshita, T. Wenger, and D.~S. Weiss, Science {\bf 305},  1125  (2004).

\bibitem{paredes04}
B. Paredes {\it et~al.}, Nature {\bf 429},  277  (2004).

\bibitem{girardeau60}
M. Girardeau, J. Math. Phys. {\bf 1},  516  (1960).

\bibitem{petrov00}
D.~S. Petrov, G.~V. Shlyapnikov, and J.~T.~M. Walraven, Phys. Rev. Lett. {\bf
  85},  3745  (2000).

\bibitem{dunjko01}
V. Dunjko, V. Lorent, and M. Olshanii, Phys. Rev. Lett. {\bf 86},  5413
  (2001).

\bibitem{alon05}
O.~E. Alon and L.~S. Cederbaum, Phys. Rev. Lett. {\bf 95},  140402  (2005).

\bibitem{hao06}
Y. Hao, Y. Zhang, J.~Q. Liang, and S. Chen, Phys. Rev. A {\bf 73},  063617
  (2006).

\bibitem{zoellner06a}
S. Z{\"o}llner, H.-D. Meyer, and P. Schmelcher, Phys. Rev. A {\bf 74},  053612
  (2006).

\bibitem{zoellner06b}
S. Z{\"o}llner, H.-D. Meyer, and P. Schmelcher, Phys. Rev. A {\bf 74},  063611
  (2006).

\bibitem{deuretzbacher06}
F. Deuretzbacher, K. Bongs, K. Sengstock, and D. Pfannkuche, Phys. Rev. A {\bf
  75},  013614  (2007).

\bibitem{girardeau07}
M.~D. Girardeau and A. Minguzzi, Phys. Rev. Lett. {\bf 99},  230402  (2007).

\bibitem{mey90:73}
H.-D. Meyer, U. Manthe, and L.~S. Cederbaum, Chem.\ Phys.\ Lett. {\bf 165},  73
   (1990).

\bibitem{bec00:1}
M.~H. Beck, A. J{\"a}ckle, G.~A. Worth, and H.-D. Meyer, Phys.\ Rep. {\bf 324},
   1  (2000).

\bibitem{zoellner07a}
S. Z{\"o}llner, H.-D. Meyer, and P. Schmelcher, Phys. Rev. A {\bf 75},  043608
  (2007).

\bibitem{zoellner07c}
S. Z{\"o}llner, H.-D. Meyer, and P. Schmelcher, Phys. Rev. Lett. {\bf 100},
  040401  (2008).

\bibitem{zoellner07b}
S. Z{\"o}llner, H.-D. Meyer, and P. Schmelcher, arXiv:0801.1090v1  .

\bibitem{kos86:223}
R. Kosloff and H. Tal-Ezer, Chem.\ Phys.\ Lett. {\bf 127},  223  (1986).

\bibitem{mey03:251}
H.-D. Meyer and G.~A. Worth, Theor.\ Chem.\ Acc. {\bf 109},  251  (2003).

\bibitem{cirone01}
M.~A. Cirone, K. G\'{o}ral, K. Rzazewski, and M. Wilkens, J. Phys. B {\bf 34},
  4571  (2001).

\bibitem{girardeau01}
M.~D. Girardeau, E.~M. Wright, and J.~M. Triscari, Phys. Rev. A {\bf 63},
  033601  (2001).

\bibitem{kolomeisky00}
E.~B. Kolomeisky, T.~J. Newman, J.~P. Straley, and X. Qi, Phys. Rev. Lett. {\bf
  85},  1146  (2000).

\bibitem{minguzzi02}
A. Minguzzi, P. Vignolo, and M.~P. Tosi, Phys. Lett. A {\bf 294},  222  (2002).

\end{thebibliography}

\end{document}